# High-temperature superconductivity in one-unit-cell FeSe films

Ziqiao Wang[1], Chaofei Liu[1], Yi Liu[1] and Jian Wang[1, 2*]

[1] International Center for Quantum Materials, School of Physics, Peking University, Beijing 100871, China
[2] Collaborative Innovation Center of Quantum Matter, Beijing, China

*E-mail: jianwangphysics@pku.edu.cn

**Abstract**
Since the dramatic interface enhancement of superconducting transition temperature ($T_c$) was reported in one unit-cell FeSe film grown on SrTiO$_3$ substrate (1-UC FeSe/STO) by molecular beam epitaxy (MBE), related research on this system has become a new frontier in condensed matter physics. In this paper, we present a brief review on this rapidly developing field, mainly focusing on the superconducting properties of 1-UC FeSe/STO. Experimental evidences for the high-temperature superconductivity in 1-UC FeSe/STO, including the direct evidences revealed by transport and diamagnetic measurements, and other evidences from scanning tunneling microscope (STM) and angle-resolved photoemission spectroscopy (ARPES), are overviewed. Potential mechanisms of the enhanced superconductivity are discussed. There are accumulating arguments suggesting that the strengthened Cooper pairing in 1-UC FeSe/STO originates from the interface effects, specifically charge transfer and coupling to phonon modes in TiO$_2$ plane. The study of superconductivity in 1-UC FeSe/STO not only sheds a new light on the mechanism of high-temperature superconductors with layered structures, but also provides the insight to explore new superconductors by interface engineering.

## 1. Introduction

The discovery of superconductivity in fluorine-doped LaFeAsO with superconducting transition temperature ($T_c$) of 26 K triggered a flurry of research activity in iron-based superconductors [1-6]. The maximum $T_c$ was soon reported up to ~ 55 K [4, 5] for bulk iron-based superconducting materials. High $T_c$ beyond the McMillan limit of the conventional superconductors, as well as distinct properties of electronic structure and pairing mechanism [7], make iron-based superconductors the second class of high-$T_c$ superconductors besides cuprate superconductors [8]. All iron-based superconductors contain iron pnictide (FeAs) [1-6, 9] or iron chalcogenide (FeSe, FeS$_x$Se$_{1-x}$ or FeTe$_x$Se$_{1-x}$) [10-48] layer, which is generally considered to be responsible for the superconducting properties and plays similar role as CuO$_2$ plane in cuprate superconductors. Although high-temperature superconductivity has been discovered for over 30 years, the mechanism is still in suspense [49-51]. The discovery of iron-based superconductors provides a new arena to understand the high-temperature superconductivity.

FeSe is a model system to investigate the superconducting properties of iron-based materials due to its simple crystal structure, clean superconducting phase and low toxicity. Although the superconducting transition temperature for bulk FeSe is relatively low ($T_c$ ~ 8 K) [24, 25], it can be significantly enhanced by applying high pressure [26-28], intercalating alkali metal atoms like potassium (K) [31-39], electric field tuning [46-48], or most amazingly, by growing ultrathin FeSe films on STO substrates by MBE [10-23]. Representative works devoted to enhance $T_c$ in FeSe and its related systems are summarized in table 1.

Superconducting films, especially that in the 2D limit, have gained great attention in both basic and applied researches. It was long believed that superconductivity in a 2D system was impossible since fluctuations would destroy long-range order even at very low temperatures, until Thouless and Kosterlitz revealed the superfluid or superconducting topological phase transition in 2D systems [52]. Some representative discoveries regarding 2D superconductivity emerge in the following aspects. (1) 2D Superconductivity may occur at the interface between two non-superconducting materials, e.g. $LaAlO_3/SrTiO_3$ interface [53]. (2) Substrates may significantly enhance the $T_c$ of the superconducting films, e.g. 1-UC FeSe/STO [10-23] and $La_2CuO_{4+\delta}/La_{1.55}Sr_{0.45}CuO_4$ [54, 55]. (3) Since fluctuations and disorder play important role in low dimensional systems [56], some new phenomena beyond the classical mean field theory emerge in superconducting films, such as quantum phase transition [57] and quantum Griffith singulartity [58]. (4) Disordered superconducting thin films close to a quantum phase transition provide a platform to detect the Higgs mode [59]. (5) Film thickness become an important parameter to tune the superconducting properties in ultrathin films even with single atomic layer change, for instance, $T_c$ oscillates with film thickness in Pb films on Si substrates [60].

Recently, the state-of-art film-growth technique in ultrahigh vacuum chamber has made the fabrication of atomically flat crystalline thin films possible. For example, high quality samples with accurate stoichiometry and few defects can be realized in a MBE chamber. Furthermore, the MBE chamber can be linked to STM or ARPES chamber in the ultrahigh vacuum environment, which enables the study of pristine surfaces without exposing the sample to the atmosphere.

Superconducting FeSe crystalline thin films can be prepared by layer-by-layer epitaxial growth on double-layer graphene along (001) crystal direction (figure 1) [61, 62]. Graphene substrates are acquired by heat-treatment of SiC (0001) substrate to evaporate silicon atoms at surface [63, 64]. One FeSe unit-cell consists of three atomic layers with central Fe layer sandwiched in two adjacent Se layers (inset of figure 1(a)). FeSe film and graphene substrate are coupled by weak van der Waals type interaction. Thus, the lattice constant of FeSe film on graphene is very close to that of bulk material (~ 3.77 Å). In addition, the FeSe films grown on graphene show lower $T_c$ with thinner thickness. Superconducting coherence peak can be clearly observed in differential tunneling conductance ($dI/dV$) spectra by scanning tunneling spectroscopy (STS). Analysis of normalized $dI/dV$ spectra at various temperatures gives $T_c$ ~ 7.8 K for 8-UC FeSe films (figure 1 (b)) and $T_c$ ~ 3.7 K for 2-UC films (figure 1 (c)). Superconductivity cannot be observed in 1-UC FeSe films down to 2.2 K. The relation between superconducting transition temperature $T_c$ and film thickness $d$, as shown in figure 1 (d), exhibits a good agreement with empirical formula $T_c(d) = T_{c0}(1 - d_c / d)$ [65], which is adoptable to both conventional superconducting films like Pb films on Si (111) substrates [66] and unconventional superconducting films like $YBa_2Cu_3O_x$ films on STO substrates [67]. The extrapolated $T_{c0}$ ~ 9.3 K is consistent with $T_c$ of bulk FeSe, suggesting graphene substrate has little effect on the superconducting properties of FeSe films.

1-UC FeSe films grown on STO substrates by MBE, however, exhibit extremely enhanced superconductivity (figure 2) [10, 68] compared to bulk FeSe and FeSe films on graphene. After epitaxial growth of 1-UC FeSe on Nb-doped (001)-orientated single crystal STO and proper annealing process, the sample was transferred into STM chamber without breaking the ultrahigh vacuum condition. An unexpectedly large superconducting-like gap of 20.1 meV was identified from the U-shaped conductance spectrum of the 1-UC FeSe/Nb: STO film (figure 2 (c)). This gap size is almost one order of magnitude larger than 2.2 meV for bulk FeSe, and it would correspond to $T_c$ ~ 80 K [10] if the ratio between the superconducting gap and $T_c$ ($2\Delta/k_B T_c$) is assumed to be similar to that of bulk FeSe. Surprisingly, tunneling spectra taken on 2-UC or thicker FeSe films do not show any sign of superconductivity (figure 2(d)). This feature is in sharp contrast to that observed in FeSe/graphene, where $T_c$ decreases while reducing film thickness (figure 1(d)) and superconductivity disappears in 1-UC films. The difference indicates that the STO substrate and interface play an essential role in the enhanced superconductivity of 1-UC FeSe/STO.

The discovery of potential high-temperature superconductivity in 1-UC FeSe/STO has opened up a new frontier for superconductivity [55]. High $T_c$ with simple structure, enhanced superconductivity by interface engineering [18, 69] and possible topological nature [70-74] make this system a highlight in condensed matter physics and material science. In this paper, we present a brief review on this rapidly developing direction. We will illustrate the high-temperature superconductivity of 1-UC FeSe/STO in section 2. The direct evidences revealed by transport and diamagnetic measurements, and other evidences from STM and ARPES, are overviewed therein. In section 3, potential mechanisms of the $T_c$ enhancement are discussed. Finally, we end with a conclusion and perspective in section 4.

## 2. Evidences for high-temperature superconductivity in one-unit-cell FeSe films on SrTiO$_3$ substrates

2.1 Transport and diamagnetic measurements

While a superconducting-like gap as large as 20 meV had been revealed in 1-UC FeSe/STO, direct evidence from electrical transport and magnetic measurements was still needed to confirm the observed superconductivity. Initial transport measurements on 5-UC FeSe film covered by 20 nm amorphous Si protection layer show a superconducting transition with resistance decrease starting at ~ 53 K [10], but evidence of exact zero resistance is absent. Therefore, systematic study on electrical transport and magnetic measurements of 1-UC FeSe/STO with clear evidence of zero resistance and diamagnetism is important [11, 75, 76].

For *ex situ* electrical transport measurements, insulating single crystal STO (001) was chosen as the substrate. STO substrates were pretreated by chemical etching with 10% HCl solution and subsequent thermal annealing at oxygen atmosphere to obtain TiO$_2$-terminated surface before transferred into ultrahigh vacuum MBE chamber. FeSe films were grown by coevaporation of Fe and Se with flux ratio of ~ 1:10 on STO substrates. After proper annealing process, 1-UC FeSe films were covered by 10-UC FeTe layer and 30 nm amorphous Si layer to protect the sample from deterioration in air [11]. Major transport properties of FeSe films were measured by standard four-electrode method (inset of figure 3(a)) in commercial physical property measurement system (Quantum Design PPMS-16). It should be mentioned that the indium electrodes used here can easily penetrate through the protection layers and connect to the nethermost FeSe layer on STO substrate.

Temperature dependence of resistance for 1-UC FeSe/STO, as shown in figure 3(a), exhibits an obvious superconducting transition with $T_c^{onset}$ ~ 40.2 K and $T_c^{zero}$ ~ 23.5 K. Here, $T_c^{onset}$ is acquired by extrapolating the normal resistance curve and the sharp resistance drop, and $T_c^{zero}$ is defined as the temperature where resistance drops below the instrumental resolution. $T_c^{onset}$ ~ 40.2 K is almost 5 times of $T_c$ ~ 8 K for bulk FeSe, demonstrating significant interface enhanced superconductivity. Gradual deviation from normal resistance starts at ~ 54.5 K, suggesting the superconductivity may survive up to this high temperature. For comparison, $R(T)$ curves for 10-UC FeTe capping layer on STO substrate with or without additional amorphous Si layer exhibit insulating like behavior. In addition, STS suggests proximity effect cannot turn STO substrate and FeTe protection layer into superconducting [11]. These facts indicate high-temperature superconductivity in FeSe/STO system is limited to single unit-cell FeSe layer with just 0.55 nm thickness.

Magnetic susceptibility measurements for the same sample by commercial magnetic property measurement system (MPMS-SQIUD-VSM) show Meissner effect and demonstrate the superconductivity in FeSe/STO system (figure 3(b)). Sharp drop in $M(T)$ curve at ~ 25 K is a clear evidence for diamagnetism corresponding to $T_c^{zero}$ obtained by transport measurements, and typical magnetic hysteresis loop measured at 2 K also exhibits superconducting characteristic [11]. Diamagnetism of 1-UC FeSe films on insulating STO substrates was further confirmed by two-coil mutual inductance measurements for another 1-UC FeSe/STO sample. Magnetic susceptibility measurements were also been carried out in 2-UC to 4-UC FeSe/STO films, giving clear evidence of diamagnetism in multilayer FeSe/STO [77].

Besides high $T_c$, the 1-UC FeSe/STO also exhibits high critical magnetic field ($B_c$) and large critical current density ($J_c$) [11]. To determine the upper critical field $B_{c2}$, magnetoresistance at various temperatures was measured by employing pulsed magnetic field up to 52 T in perpendicular (figure 4(a)) and parallel (figure 4(b)) orientations. In perpendicular field, the sample resistance at 1.4 K remains at zero until the field increases above 30 T and is still much smaller than the normal state resistance even at 52 T. When magnetic field is applied parallel to the film, $B_c$ is even higher than that in perpendicular field, reflecting the 2D nature of superconductivity in 1-UC FeSe/STO. Moreover, 1-UC FeSe film exhibits an unexpected large critical current density ($J_c$). The critical current at 2 K reaches 13.3 mA, corresponding to $J_c$ ~ 1.7 × $10^6$ A/cm$^2$, which is two orders of magnitude higher than that of bulk FeSe ($J_c$ ~ 2.2 × $10^4$ A/cm$^2$) [78].

2D superconducting property is further supported by the observation of Berezinski-Kosterlitz-Thouless (BKT) like transition (figure 5) [11], which is a topological phase transition. Temperature dependence of the power-law exponent $\alpha$ ($V \propto I^\alpha$) indicates BKT-like transition temperature $T_{BKT}$ ~ 23.1 K. BKT-like transition temperature can also be determined by using Halperin-Nelson equation $R(T) = R_0 \exp[-b/(T-T_{BKT})^{1/2}]$ [79], yielding $T_{BKT}$ ~ 23.0 K, consistent with the result from the $V(I)$ curves.

The resistive transition of high-$T_c$ superconductors in the presence of magnetic field is an active research topic [19, 80, 81]. $R_{sq}(T)$ curves and corresponding $\ln[R_{sq}]$-(1/$T$) curves for 1-UC FeSe/STO at various magnetic fields are shown in figure 6(a) and 6(b). The resistance transition at low temperature region can be well described by the theory of thermally activated flux flow using the Arrhenius relation $R(T,H) = R(H)\exp[-U_0(H)/T]$, where $U_0$ is the activation energy. The slopes obtained from linear fittings (pink lines in figure 6(b)) give the activation energy $U_0(H)$ at various magnetic fields (figure 6(c)). Extrapolation of these fitting lines has a common intersection point at $T_m$ ~ 38 K, which is close to $T_c^{mid}$ (defined as the temperature where resistance drops to half of normal resistance) from $R_{sq}(T)$ curves. Power-law fittings $U_0(H) \sim H^{-\gamma}$ are performed on $U_0(H)$ data, showing a crossover from $\gamma$ ~ 0.14 for $\mu_0 H$ < 3.4 T to $\gamma$ ~ 0.60 for $\mu_0 H$ > 3.4 T. The noteworthy change of $\gamma$ marks a crossover from a single-vortex pinning dominated regime to a collective flux creep regime [19]. Similarly, a recent work analyzed the activation energy $U_0(J)$ by measuring $R_{sq}(T)$ curves at various current densities $J$ [80]. $U_0(J)$ saturates to a finite value when $J \rightarrow 0$, suggesting that the long-range vortex lattice correlations may be absent in 1-UC FeSe/STO possibly due to the strong thermal fluctuations in 2D materials [80].

Hall measurements were carried out in 1-UC FeSe films to determine carrier type and carrier density [19]. Hall resistance $R_{xy}(H)$ curves at various temperatures exhibit good linear relation (figure 7(a)), where the influence of FeTe protection layers has been subtracted. Hall coefficient and carrier density derived from $R_{xy}(H)$ curves (figure 7(b)) indicate the dominant carriers transform from hole type at high temperatures to electron type at low temperatures. Electron carrier density in low temperature regime is in the order of $10^{15}/cm^2$. The high carrier density is generally believed to be a necessary factor to achieve superconductivity. Similar dominant carriers transformation from hole type at high temperatures to electron type at low temperatures also happens in 2-UC FeSe/STO [19].

Transport measurements of multilayer FeSe/STO with different film thicknesses and annealing conditions have been studied [19, 82]. As mentioned before, the indium electrodes can penetrate to the nethermost FeSe layer on STO substrate. Therefore observation of superconductivity in multilayer FeSe film by transport measurement is not contradictory to the fact that superconducting gap is absent on the top surface of multilayer film by STM or ARPES measurement [10, 14, 83] if the nethermost FeSe layer is superconducting in multilayer FeSe film. FeSe/STO system exhibits unique superconducting property, i.e. onset $T_c$ decreases with increasing film thickness (figure 8(a) and 8(b)) [82], which is opposite to that observed in traditional superconducting films, including FeSe films grown on graphene [61]. Besides, proper annealing process is crucial to the observation of superconductivity in FeSe/STO system. While as-grown 5-UC FeSe film is insulating, annealing at 500 °C for 36 h makes it a superconductor with $T_c^{onset}$ ~ 39 K (figure 8(c)). Excessive annealing, however, would make $T_c$ lower, possibly due to over doping or gradual evaporation of FeSe at high annealing temperature [83]. Annealing process is accompanied by transformation of dominant charge carrier from hole to electron at low temperatures (figure 8(d)), indicating effective electron doping from STO substrate during annealing process.

As a conclusion of this section, observation of zero resistance and diamagnetism unambiguously demonstrate superconductivity in 1-UC FeSe/STO. However, in most cases, $T_c$ by *ex situ* transport or magnetic measurement is not as high as the expectation from STM and ARPES measurements (see table 1). This result can be understood in the following aspects. First, superconductivity in 1-UC FeSe film is dependent on the level of electron doping from STO substrates [13, 15, 16, 83]. For *ex situ* transport measurements of multilayer FeSe films or 1-UC FeSe films with FeTe capping, carriers from STO substrates may spread over FeSe layers in multilayer FeSe films or spread to FeTe protection layers in 1-UC FeSe films with FeTe capping, making the doping level of the interfacial FeSe layer lower than the situation with only 1-UC FeSe layer on STO [83], which is the situation for STM and ARPES studies. Second, non-superconducting FeTe layer has different magnetic structure with FeSe layer [84-88], and this difference may have negative effect on the superconductivity of 1-UC FeSe/STO. Third, the doping level in 1-UC FeSe/STO depends on the annealing temperature and lasting time [13, 15]. STS suggests that a high annealing temperature (500 ℃ or above) is helpful to form a large superconducting gap [15]. However, FeSe would evaporate gradually during annealing process at such high temperature [83]. Since keeping FeSe film well connected is necessary for *ex situ* transport measurements, the high annealing temperature may decrease $T_c$ obtained by transport measurements. Last but not least, exposing to air is damaging to ultrathin FeSe films even though protection layer is covered.

It is worthy to mention that *in situ* special four-tip electrical transport measurement technique was developed (figure 9(a)) to investigate the superconductivity in 1-UC FeSe/Nb: STO films without exposing the sample to air [20, 89]. As shown in figure 9(b), zero resistance disappears at a very high temperature of ~ 109 K even beyond the expectation from STM and ARPES measurements. Although the result is very exciting, the transport measurement is not standard four-electrode method and the substrate is conducting. Thus, further confirmation from other groups and techniques is necessary. The indication of $T_c$ above liquid nitrogen temperature was also revealed by *ex situ* diamagnetic measurements of 1-UC FeSe/Nb: STO films [19, 55]. Magnetic susceptibility $M(T)$ curves of 1-UC FeSe film after subtracting the influence of STO substrate and FeTe capping layer exhibit a gradual decrease in magnetization starting from ~ 85 K (figure 10) [19]. Besides, some two-coil mutual inductance measurements of 1-UC FeSe film with a capping layer of 2-UC (Fe$_{0.96}$Co$_{0.04}$)Se/2-UC FeSe/Se also showed a formation of diamagnetic screening up to ~ 65 K [21].

2.2 STM probing

Scanning tunneling microscopy/spectroscopy (STM/STS) is a powerful technique for the characterization of superconducting FeSe/STO system [10, 15, 22, 23, 90-94]. MBE-STM combined system in ultrahigh vacuum environment makes it possible to investigate the intrinsic physical properties of extremely clean surface with very few impurity adatoms. Surface topography with atomic resolution can be presented by STM image, and local density of state (LDOS) near Fermi level ($E_F$) can be probed through tunneling conductance spectra (d$I$/d$V$) [95]. STS can thus provide the evidence of superconductivity by studying the size of energy gap and its evolution with temperature and magnetic field in a superconductor. In addition, by Fourier transforming the real space d$I$/d$V$ mapping, the *k*-space information of the sample can be accessed [91, 92, 96]. Another advantage of STM/STS is the capability of detecting local response of superconductivity to magnetic or non-magnetic impurities [91, 97-99]. Characteristics of in-gap impurity states provide information on the pairing symmetry of superconductivity [91, 97], and particularly zero-energy in-gap states may correspond to Majorana bound states localized at the edge of topological superconductors [98, 99].

STM/STS results have shown significant effect of annealing process on surface morphology and superconducting gap of 1-UC FeSe/STO [15, 90]. As-grown FeSe film appears quite rough and exhibits an asymmetric semiconducting gap. With increasing annealing temperature, the FeSe film becomes smoother, and transforms from semiconducting to metallic and finally to superconducting state. Figure 11(a) shows the evolution of superconducting gap with increasing annealing temperature at 450, 480, 500, 510, and 530 ℃, respectively. Each annealing stage lasts for 2 hours. The superconducting gap gradually becomes larger with increasing annealing temperature, and reaches ~ 15.4 meV after annealing at 530 ℃ (figure 11(b)). Figure 11(c) displays a series of normalized d$I$/d$V$ spectra taken at various temperatures on FeSe surface after annealing at 530 ℃. The coherence peaks are gradually suppressed and the zero bias conductance (ZBC) continuously increases with increasing temperature (figure 11(d)). ZBC shows a linear dependence on temperature, suggesting an extrapolated $T_c$ ~ 68 K. Though grown on insulating STO substrates, the $T_c$ value is close to that estimated for 1-UC FeSe/Nb: STO. We know the pre-annealing process of STO substrate in ultrahigh vacuum environment induces 2D electron gas (2DEG) at the surface [15, 100], which becomes carrier reservoir and transfers electrons to FeSe layer in post-annealing process, as illustrated by transport [11, 82] and ARPES [13, 16] measurements. Surprisingly, the formation of highly metallic 2DEG at the surface of STO is independent on bulk carrier densities over a wide range from less than $10^{13}$ cm$^{-3}$ (insulating STO) to $10^{20}$ cm$^{-3}$ (Nb: STO) [101]. It explains why the $T_c$ for 1-UC FeSe films on insulating and conducting STO could be close.

By measuring the local response of superconductivity to impurities and through quasi-particle interference (QPI) patterns, STM/STS has provided the evidence of plain $s$-wave pairing symmetry in 1-UC FeSe/STO system [91]. QPI patterns (figure 12), acquired by Fourier transform of the real space d$I$/d$V$ mapping, reflect scatterings between and within the electron pockets. Annealed 1-UC FeSe/STO has four ellipselike electron pockets at M points. Possible scattering channels ($q_1$, $q_2$, $q_3$) between and within these electron pockets (figure 12(a)) correspond to three scattering rings in QPI pattern (figure 12(c)). Integrated intensities over different scattering rings have similar energy dependences near the gap edge (figure 12(d)), suggesting that none of these scattering channels has significant differences and the signs of superconducting gap $\Delta_k$ on different M points are similar. This observation is consistent with the plain $s$-wave or $s_\pm$-wave pairing (± refers to the sign change of the superconducting order parameters between electron and hole Fermi surfaces) but incompatible with $d$-wave pairing. Same conclusion can be also deduced from the magnetic field dependence of QPI (figure 12(e, f)). Probing the response of superconductivity to local impurities (figure 13) is another common way to determine the pairing symmetry. While magnetic impurities (Cr and Mn) suppress the superconductivity and induce in-gap states, the superconducting gap is basically undisturbed by non-magnetic impurities (Zn, Ag and K). Since $d$ or $s_\pm$ pairing is sensitive to non-magnetic impurities [102, 103], plain $s$-wave pairing is the most likely pairing symmetry in 1-UC FeSe/STO.

2.3 ARPES measurements

Angle-resolved photoemission spectroscopy (ARPES) provides another approach to examine the superconductivity in FeSe/STO system [12-14, 16-18, 104]. Reminiscent of STM study, the superconducting transition temperature $T_c$ can be extracted from photoemission spectra (energy distribution curves, EDCs) by measuring the gap-opening temperature and temperature dependence of the gap. A superconducting signature with $T_c$ ~ 65 K [13, 14] has been revealed by ARPES in 1-UC FeSe/STO after appropriate post-annealing procedures. Over other techniques, ARPES provides direct information on band structures and the superconducting gap can be measured at specific $k$ points in momentum space, which shed light on the investigations of pairing symmetry and pairing mechanisms of the superconductivity.

Figure 14 shows the Fermi surface and band structure of 1-UC FeSe/STO. The Fermi surface consists only of electron pockets around M points (figure 14(a)). In particular, holelike Fermi surface at the Brillouin zone center (Γ), which exists in most iron-based superconductors including iron pnictides [105] and bulk FeSe [106], is absent in 1-UC FeSe/STO. The lack of hole pocket implies that 1-UC FeSe/STO film is electron doped. In fact, similar feature has been observed in electron-doped iron chalcogenides, such as $K_xFe_{2-y}Se_2$ [35, 107]. The absence of hole Fermi surface means $s_\pm$ pairing scenario [108] may not be a proper description of the superconductivity in 1-UC FeSe/STO. Band structures along cut 1 and cut 2 are shown in figure 14(b), exhibiting hole and electron band, respectively. The holelike band at Γ point sinks ~ 80 meV below the Fermi level, explaining the absence of hole Fermi surface. High-resolution Fermi surface mapping presents the detailed structure of electron Fermi surface [109]. Two overlapping ellipselike electron pockets ($\delta_1$, $\delta_2$) are resolved at M point, though the photoemission intensity of $\delta_2$ pocket is larger than that of $\delta_1$ pockets (figure 14(c, d)).

The superconducting transition temperature $T_c$ was acquired by analysing the temperature dependence of the symmetrized photoemission spectra along the $\gamma$ Fermi surface (figure 15(a)). The superconducting-like gap shrinks with increasing temperature and disappears at ~ 65 K for well annealed 1-UC FeSe/STO samples [13]. The evolution of superconducting gap with temperature exhibits good agreement with BCS fitting (figure 15(b)), giving $T_c$ ~ 65 K and gap size of ~ 19 meV consistent with the STS results [10]. As for the gap distribution in the momentum space, anisotropic but nodeless superconducting gap (figure 15(c)) was revealed by a recent high resolution ARPES experiment [109]. Gap maxima ~ 12 meV locate along the major axis of the ellipse, while gap minima ~ 8 meV locate at the intersection of two ellipselike electron pockets. Besides, pronounced anisotropy of the gap size with over 50% variation (range from 8.5 meV to 17.2 meV) was observed in 1-UC FeSe films on Nb: $STO/KTaO_3$ with larger lattice constant due to the tensile strain from substrate [18].

2.4 Evidences from other techniques

Some other techniques have also brought new insight into the detection of superconductivity in 1-UC FeSe/STO system. For example, superconducting transition with $T_c$ ~ 68 K and superconducting gap of ~ 20.2 meV were observed in single layer FeSe/STO system by using ultrafast spectroscopy [110]. Especially, the electron-phonon coupling strength $\lambda \approx 0.48$ can be obtained by measuring the lifetime of the non-equilibrium quasiparticle pumped by femtosecond laser. Such information is inaccessible by most of other methods and of importance in understanding the high-$T_c$ superconductivity in FeSe/STO system.

In addition, the superconductivity at 62 K was reported in FeSe/STO by transverse field muon spin rotation and relaxation (TF-μSR) spectra [111]. Measurements of the field distribution in the vortex state give the effective penetration depth $\lambda$ and hence the superfluid density $n_s \propto \lambda^{-2}$. The density of paired electrons is estimated to be $n_s$ ~ $6.08 \times 10^{21}$ $cm^{-3}$, consistent with the doping level revealed by ARPES (~ 0.12 electron per Fe corresponds to $n_s$ ~ $6 \times 10^{21}$ $cm^{-3}$) [13] and transport measurements ($n_s^{2D}$ ~ $10^{15}$ $cm^{-2}$, or $n_s$ ~ $10^{22}$ $cm^{-3}$) [19]. The temperature dependence of $n_s$ is in good agreement with BCS fitting, suggesting an *s*-wave superconducting state with gap size of ~ 10.2 meV. Besides, polarized muons probe does not detect the indication of magnetism and thus support the absence of SDW order in single-layer FeSe.

**3. Discussion on the mechanism of interface enhanced superconductivity in 1-UC**

**FeSe/STO**

Experimental discoveries of high-temperature superconductivity in 1-UC FeSe/STO have triggered great interest in the community of condensed matter physics to figure out the underlying mechanism. Now it is widely accepted that the dramatic enhancement of $T_c$ originates from the FeSe/STO interface. Several factors regarding the interface have been considered, for instance, the tensile strain due to in-plane lattice constant difference between FeSe film and STO substrate, the electron doping from the STO substrate, and the electron-phonon coupling between the electrons in FeSe and the high-frequency phonons in STO, etc.

The role of tensile strain on the enhancement of $T_c$ was studied by growing 1-UC FeSe films on different substrates with various in-plane lattice constants [18, 69, 112]. For comparison, the in-plane lattice constant, the tensile strain and corresponding $T_c$ are listed in table 2. Although $T_c$ seems to have positive correlation with the tensile strain in 1-UC FeSe film, the effect of tensile strain is not large enough to explain the giant enhancement of $T_c$. Particularly, $T_c$ is still as high as ~ 70 K in 1-UC FeSe/Nb: $BaTiO_3$/$KTaO_3$ (rotated domain), although the tensile strain can be negligible. Moreover, recent STM measurements reveal that nearly strain-free FeSe films on anatase $TiO_2$ (001) exhibit large superconducting gap of similar size to 1-UC FeSe/STO [113]. DFT calculations show that the magnetic interaction is sensitive to the lattice constant, e.g. the next-nearest-neighbor coupling parameter $J_2$ increases about 40% when the lattice constant increases just a few percent [114]. Since in reality the tensile strain only has a small effect (if it does have some effect) on $T_c$, magnetic interaction mediated pairing mechanism alone cannot unveil the secret of high $T_c$ in FeSe/STO system.

There are accumulating arguments suggesting electron doping and electron-phonon coupling are responsible for raising $T_c$ in 1-UC FeSe/STO system. High-temperature superconductivity with $T_c$ ~ 40 K has been reported in K-coated [40] or liquid gated [46-48] FeSe films, indicating that high $T_c$ can be achieved in FeSe layer once it is optimally doped. Observation of replica bands by ARPES [17, 122, 123] reveals the crucial role of interfacial coupling between the FeSe electrons and phonon modes in STO. These findings imply FeSe/STO system is likely to be understood in conventional BCS model [115].

3.1 The role of electron doping

Carrier density is a primary factor for superconductivity. In BCS model, $T_c$ is roughly estimated as $T_c \sim \theta_D e^{-1/N(0)V}$ in the weak-coupling limit [116], where $\theta_D$, $N(0)$ and $V$ represent the Debye temperature, the density of state (DOS) at the Fermi energy and the phonon-mediated attractive interaction, respectively. Clearly, in this conventional scenario, superconductivity is not favorable in band insulators with low carrier density since $T_c \to 0$ in the $N(0) \to 0$ limit. However, specific band insulators like STO [117] and $MoS_2$ [118], can be tuned into superconducting state by increasing the carrier density. Some common carrier density tuning methods include chemical doping (e.g. Nb-doped STO [117], alkali metal dopants intercalated FeSe [31-39] or $MoS_2$ [118], K-coated FeSe [40-45]), electric field tuning (e.g. liquid gated FeSe [46-48] or $MoS_2$ [118]), thermal treatment (e.g. STO turns from insulating to conductive after heating above 800 ℃ in oxygen-free environment [101, 117]), and charge transfer at interface (e.g. 1-UC FeSe/STO [10-23]). The dependence of $T_c$ with dopant concentration $x$ in both cuprates [119] and iron pnictides [9] exhibits dome-like shape. The doping of a Mott insulator is generally viewed as a starting point to understand the physics of high-temperature superconductivity in cuprates [119], though the exact pairing mechanism is still under debate.

We have shown in the previous section that post-annealing process is important to the superconductivity in FeSe/STO system. In fact, both the transport and STM measurements reveal the insulating-like behavior in as-grown FeSe films on STO. Subsequent annealing processes induce insulator-superconductor transition, and within a certain range, $T_c$ increases with increasing annealing temperature [15, 82]. Post-annealing process is accompanied by electron doping, which has been demonstrated by both ARPES and transport measurements [16, 82]. ARPES results show that the holelike pocket at Γ point gradually sinks below the Fermi energy and at the same time the electron Fermi surface at M point gradually enlarges, indicating electron concentration in 1-UC FeSe/STO increases continuously during the annealing process [13, 16]. Hall measurements also show that after annealing $n$ type carriers dominate in FeSe/STO above $T_c$ [19, 82]. However, excessive annealing at relatively high temperature would deteriorate the superconductivity, for instance, 5-UC FeSe/STO after 55 h annealing at ~ 500 °C exhibits lower $T_c$ than that annealing for 36 h (figure 8(c)) as revealed by transport measurements [82]. This dome-like behavior is reminiscent of the cuprate superconductors and probably results from over doping or gradual evaporation of FeSe at high annealing temperature.

Recently, electrical double-layer transistor (EDLT) technique has been used to electrostatically dope carriers to FeSe films on different substrates [46-48]. Figure 16 displays the electric field induced superconductivity in electrochemically etched ultrathin FeSe films on STO and MgO substrates [46]. Electric field has a significant effect on FeSe films, especially for those with thickness below 10 UC (figure 16(c, d)). For a sample with 3.7 nm thick, onset $T_c$ increases with increasing gate voltage and reaches around 40 K at $V_G = 5$ V (figure 16(e)). $T_c$ enhancement from ~ 8 K (the bulk value) to ~ 40 K suggests that electron doping contributes to the observed high-$T_c$ superconductivity in FeSe layer.

Another noteworthy example is the observation of superconductivity with $T_c$ ~ 48 K in K-coated 3-UC FeSe/STO [40]. As mentioned before, the superconductivity in FeSe/STO system is limited in the nethermost FeSe layer [10, 11] and it is very hard to tune the upper layer into heavily electron-doped region by substrates [83]. Thus the high carrier density at the surface of 3-UC FeSe/STO is provided by the deposited K atoms, similar to the situation in alkali metal dopants intercalated FeSe [31-39]. All these electron-doped iron chalcogenides (1-UC FeSe/STO [12-14], K-coated 3-UC FeSe/STO [40], K-coated bulk FeSe [44, 45, 120] and $K_xFe_{2-y}Se_2$ [35]), whose Fermi surfaces consist only of electron pockets, exhibit much higher $T_c$ than bulk FeSe. Therefore electron doping can at least partly account for the $T_c$ enhancement in FeSe/STO.

On the other hand, $T_c$ for K-coated 3-UC FeSe/STO (~ 48 K), K-coated bulk FeSe (~ 25 K) and $K_xFe_{2-y}Se_2$ (~ 30 K) is not as high as the value of 1-UC FeSe/STO (~ 65 K, as revealed by ARPES and STM), indicating that STO substrate plays distinct role in realizing high-$T_c$ superconductivity beyond merely doping electrons to FeSe layer. By systematic STM study in K-coated FeSe films on STO and graphene substrates [41-43], it is found that while the superconducting gap remains at ~ 9 meV for films on graphene by K coating, it increases from ~ 9 meV to ~ 15 meV for films on STO when the thickness is reduced to a few unit cells (figure 17(a)). The FeSe/STO interface enhancement of the superconducting gap decays exponentially with a characteristic length of 2.4 UC as film thickness increases [43], which matches well with the decay behavior of the penetrating field intensity of Fuchs-Kliewer phonon modes in STO (figure 17(b)) [121]. However, there is almost no such interface effect between FeSe and graphene [43, 61, 62]. These results point to interfacial electron-phonon coupling as the likely origin of the high-$T_c$ superconductivity in 1-UC FeSe/STO system.

3.2 Interface-enhanced electron-phonon coupling

In BCS model, electrons form Cooper pairs via phonon mediated attractive interaction [116], where the electron-phonon coupling strength dominates in determining the superconducting transition temperature. Interface-enhanced electron-phonon coupling has been suggested as a feasible origin of the dramatic $T_c$ enhancement in FeSe/STO system [14, 121-128].

The electron-phonon coupling strength is quantified as frequency-dependent electron-phonon coupling constant $\lambda(\omega)$ and Eliashberg spectral function $\alpha^2F(\omega)$, obeying

$\lambda(\omega)=2\int_0^\omega d\nu \left[\alpha^2F(\nu)/\nu\right]$ [129]. The calculated $\lambda(\omega)$ and $\alpha^2F(\omega)$ spectra (figure 18(a)) reflect stronger electron-phonon coupling strength in FeSe/STO than bulk FeSe, signaling interface-enhanced electron-phonon coupling at FeSe/STO interface [125]. Since both the interpocket and intrapocket electron-phonon coupling are attractive and hence beneficial to the even-sign $s_{++}$ pairing in FeSe/STO, $T_c$ is enhanced with $\lambda_{\text{intra}} > 0$ and $\lambda_{\text{inter}} > 0$ (figure 18(b)) [124]. Strengthened pairing might originate from the interfacial interaction between FeSe electrons and ferroelectric phonons involving the relative displacement of Ti and O atoms. Interestingly, ferroelectric transition near the gap-opening temperature has been detected in 1-UC FeSe/STO by mutual inductance and Raman spectroscopy measurements [130]. The coincidence of the ferroelectric transition and the superconducting transition temperature suggests the crucial role of ferroelectric phonon in promoting pairing in FeSe/STO system.

The detection of replica bands in 1-UC FeSe/Nb: STO (001) by high-resolution ARPES provides experimental indication for interfacial coupling between FeSe electrons and phonon modes in STO [17, 128]. As shown in figure 19(a), band structures of 1-UC FeSe along a high-symmetry cut centered at M point demonstrate an electron-like band with its bottom 60 meV below Fermi energy. Surprisingly, two extra replica bands (A' and B') of the main bands are observed. Except an energy shift of ~ 100 meV, all characteristics of the replica bands are basically identical to their corresponding main bands. Similar findings are reproduced by raw EDCs (figure 19(b)). The temperature evolution of M spectrum for 1-UC film indicates that the replica bands are robust and persist to temperatures far above the gap-opening temperature (figure 19(c)). Moreover, the replica bands only exist in 1-UC FeSe film and disappear in 2-UC or thicker film (figure 19(d)). This feature excludes the possibility of quantum well states [60, 131] as the cause of the replica bands. Instead, the energy separation of ~ 100 meV is identified as the coupling to phonon modes in STO. Since superconductivity and replica bands are both detected only within the bottom FeSe layer on STO, the coupling between FeSe electrons and STO phonons might be responsible for the enhanced Cooper pairing at FeSe/STO interface.

By assuming both electron and hole bands couple to phonon modes with energy ~ 80 meV and optimizing the coupling strength and forward-scattering parameters, the typical replica bands acquired by high-statistics scan at low temperature (figure 19(e)) can be well simulated (figure 19(f, g)). The electron-phonon coupling constant is estimated as $\lambda \approx 0.5$, close to the value ($\lambda \approx 0.48$) obtained by ultrafast optical spectroscopy [107]. Theoretical analysis points out that the phonon-mediated attraction can effectively enhance the magnetic interaction mediated Cooper pairing [132]. Specifically, $T_c$ enhancement factor $T_c(v_{eff})/T_c(0)$ is determined directly proportional to the ratio between phonon-mediated attraction strength $v_{eff}$ and antiferromagnetic exchange constant $J$ (figure 19(h)). A conservative estimate gives $T_c(v_{eff})/T_c(0) \sim 1.5$. Considering that $T_c$ for electron-doped FeSe systems without interface effect is at the level of 30-40 K (see table 1), this enhancement factor yields fairly good agreement with the gap-opening temperature. Whether the coupling to STO phonon modes has beneficial contribution to the Cooper pairing depends on the superconducting pairing symmetry. In the case of FeSe/STO with two electron pockets at M and the sign-unchanging $s$-wave pairing (supported by STM [91], ARPES [12, 17, 109] and theory [17, 124]), the total contribution of electron-phonon coupling to pairing $\lambda_{ph} > 0$ and thus enhances $T_c$. Therefore, interface-enhanced electron-phonon coupling is experimentally substantiated as an interpretation of the dramatic $T_c$ enhancement in 1-UC FeSe/STO.

The essential role of interfacial electron-phonon coupling in raising $T_c$ is further verified by a recent comparison experiment on several FeSe-based systems [122]. In general, FeSe-based superconductors can be categorized into three classes, intrinsic FeSe systems (such as FeSe films on graphene, bulk FeSe and $FeTe_xSe_{1-x}$), electron-doped FeSe systems (including K-doped multilayer FeSe/STO, K-dosed bulk FeSe and $K_xFe_{2-y}Se_2$), and electron-doped FeSe systems with interfacial coupling (including 1-UC FeSe films on STO and $TiO_2$). Figure 20 presents the Fermi surface and the band structures at M points of several representative FeSe systems of these three classes. The intrinsic FeSe systems exhibit hole pockets around Γ point [14, 106, 133] and $T_c < 20$ K. Electron-doped FeSe systems share similar band structures, i.e. no hole pockets at Γ [12-14, 35, 40, 44] and hole bands at M points are ~ 45 meV below the electron bands (figure 20(f-h)). $T_c$ for electron-doped FeSe systems without interface effects is enhanced to the level of 30-40 K, which is significantly higher than intrinsic FeSe systems. $T_c$ as high as 60 K, however, can be only detected in FeSe systems with both charge transfer and additional interfacial electron-phonon coupling, e.g. 1-UC FeSe films on STO and $TiO_2$, in which replica bands are observed (figure 20(g, h)). Coincidental detection of high $T_c$ value and replica bands signifies the importance of interfacial coupling for $T_c$ enhancement. In addition, the huge differences in lattice constant and dielectric constant between $TiO_2$ and STO ($C_4$ symmetry is broken in 1-UC FeSe/$TiO_2$ (100) with a = 3.53 Å and b = 3.95 Å, while $C_4$ symmetry is preserved in 1-UC FeSe/STO (001); and $TiO_2$ at low temperature has a much smaller dielectric constant [122]) suggests tensile strain and dielectric constant are likely unimportant for enhancing $T_c$ in these systems.

As a conclusion of this section, 1-UC FeSe films grown on various $TiO_2$-terminated substrates (STO (001) [10-21], STO (110) [22, 123], $BaTiO_3$ (001) [69], $TiO_2$ (001) [113], $TiO_2$ (100) [122]) with different in-plane lattice constants and dielectric constants exhibit similar $T_c$ value (~ 60-70 K), well above the value (~ 30-40 K) for electron-doped FeSe systems without interfacial enhancement. This fact clearly demonstrates that the interface effects, specifically charge transfer and coupling to phonon modes in $TiO_2$ plane, play critical roles in the high-temperature superconductivity of FeSe/STO. $T_c$ enhancement has been shown to depend on the electron-phonon coupling strength, though quantitative inconsistency between experimental value and theoretical calculation still exists [126]. Further investigations are necessary to better understand the interface high-temperature superconductivity observed in FeSe/STO.

## 4. Summary and perspective

We briefly review the experimental progresses and the physical understandings of the high-temperature superconductivity in 1-UC FeSe/STO and related systems. Some indications show that $T_c$ could be higher than the liquid nitrogen temperature [19, 20]. However, strong evidences are still lacking and highly desired. Since 1-UC FeSe/STO films need to be taken out from the ultrahigh vacuum chamber for *ex situ* measurements, the search for high quality protection layer that does not reduce $T_c$ is urgently encouraged. With this advancement, FeSe/STO high-$T_c$ interface superconductors would have tremendous potential for superconducting electronic devices.

Now the interface effects, specifically charge transfer and coupling to phonon modes in $TiO_2$ plane, have been identified as the most likely origins of the enhanced superconductivity, although some other mechanisms cannot be completely excluded. Research on FeSe/STO not only sheds light on the mechanism of high-temperature superconductors (cuprates and iron-based superconductors) with similar layered structure (e.g. CuO/SrO in BSCCO and FeAs/LaO in LaOFeAs) [10, 11], but also provides insight to explore new superconductors by interface engineering.

Last but not least, the potential topological nature in FeSe/STO system has gained special attention [70-74]. Dirac cone like structures have been revealed in FeSe thin films by band structure calculations [70-72] as well as ARPES measurements [73]. Furthermore, a recent study predicts the existence of 1D topological edge states at the grain boundary in 1-UC FeSe/STO, which is confirmed by STM probing [74]. These experimental evidences and theoretical supports inspire people to consider the possibility of realizing Majorana fermions in systems based on FeSe/STO or other similar structures. The realization of high-$T_c$ topological superconductors would significantly promote the development of topological quantum computation in future.

## Acknowledgments

We thank Q Y Wang, Y Xing, Y Sun, W H Zhang, L L Wang, X C Ma, Q K Xue and other collaborators in this field. This work was financially supported by the National Basic Research Programme of China (Grant Nos. 2013CB934600), the Research Fund for the Doctoral Programme of Higher Education (RFDP) of China, the Open Research Fund Program of the State Key Laboratory of Low-Dimensional Quantum Physics, Tsinghua University under Grant No. KF201501, and the Open Project Programme of the Pulsed High Magnetic Field Facility (Grant No. PHMFF2015002), Huazhong University of Science and Technology.

Tables and Figures

**Table 1. Superconductivity in FeSe and its related systems.** Representative works devoted to enhance superconducting transition temperature $T_c$ in FeSe-related systems and their methods to characterize $T_c$ are summerized in this table.

| Systems | $T_c$ | Methods to characterize $T_c$ | Refs |
|---|---|---|---|
| Bulk PbO-type β-FeSe | $T_c \sim 8$ K | Electrical transport, diamagnetic and specific heat measurements | 24, 25 |
| Bulk FeSe at high pressure | $T_c^{onset} \sim 36.7$ K, $T_c^{zero} \sim 20$ K at 8.9 GPa | Electrical transport measurements | 26 |
| | $T_c^{zero} \sim 38$ K at 6.3 GPa | Electrical transport measurements | 28 |
| Bulk FeTe$_x$Se$_{1-x}$ | $T_c^{onset} \sim 14$ K at $x = 0.6$ | Electrical transport and diamagnetic measurements | 29, 30 |
| A$_x$Fe$_{2-y}$Se$_2$ (A = Li, Na, K, Rb, Cs, Ca, Sr, Ba...) | $T_c^{onset} \sim 30.1$ K, $T_c^{zero} \sim 27.2$ K for K$_{0.8}$Fe$_2$Se$_2$ | Electrical transport and diamagnetic measurements | 31-34 |
| | $T_c \sim 30$ K, superconducting gap of $\sim$ 10.3 meV for K$_{0.8}$Fe$_2$Se$_2$ | ARPES | 35 |
| | Superconducting gap of $\sim 4$ meV for KFe$_2$Se$_2$ | STM/STS | 36 |
| A$_x$Fe$_{2-y}$Se$_2$ at high pressure | $T_c^{onset} \sim 48$ K at 12.4 Gpa for Tl$_{0.6}$Rb$_{0.4}$Fe$_{1.67}$Se$_2$ | Electrical transport and diamagnetic measurements | 37 |
| (Li$_{0.8}$Fe$_{0.2}$)OHFeSe | $T_c \sim 40$ K | Diamagnetic measurements | 38 |
| 1-UC FeSe films on Nb-doped STO (001) | Superconducting gap of $\sim 20$ meV | STM/STS | 10 |
| | $T_c \sim 65$ K | ARPES | 12-14 |
| | $T_c > 100$ K | *In situ* special four-tip electrical transport measurements | 20 |
| | $T_c \sim 85$ K (susceptibility drop starts at $\sim 85$ K) | Diamagnetic measurements (SQUID) | 19 |
| | $T_c \sim 65$ K | Diamagnetic measurements (two-coil mutual inductance measurements) | 21 |
| 1-UC FeSe films on insulating STO (001) (For ex situ transport and magnetic measurements, samples were covered by 10-UC FeTe layer.) | $T_c^{onset} > 40$ K, $T_c^{zero} \sim 23.5$ K | *Ex situ* electrical transport measurements | 11 |
| | $T_c \sim 21$ K | Diamagnetic measurements (two-coil mutual inductance measurements) | 11 |
| | $T_c \sim 25$ K | Diamagnetic measurements (SQUID) | 11 |
| | $T_c \sim 68$ K, superconducting gap of $\sim 15.4$ meV | STM/STS | 15 |
| 1-UC FeSe films on STO (110) | Superconducting gap of $\sim 17$ meV | STM/STS | 22 |
| | $T_c^{onset} \sim 31.6$ K, $T_c^{zero} \sim 16$ K | *Ex situ* electrical transport measurements | 22 |
| 1-UC FeTe$_x$Se$_{1-x}$ films on STO (001) | Superconducting gap of $\sim 16.5$ meV | STM/STS | 23 |
| | $T_c^{onset} \sim 40$ K, $T_c^{zero} \sim 21$ K | *Ex situ* electrical transport measurements | 23 |
| 1-UC FeSe films on Nb: STO/KTaO$_3$ | $T_c \sim 70$ K | ARPES | 18 |
| K-coated multilayer FeSe films on STO | $T_c \sim 48$ K for 3-UC films | ARPES | 40 |
| | Superconducting gap of $\sim 13.1$ meV for 3-UC films | STM/STS | 41, 43 |
| K-coated multilayer FeSe films on graphitized SiC (0001) | Superconducting gap of $\sim 9$ meV | STM/STS | 42, 43 |
| K-dosed bulk FeSe/ FeSe$_{0.93}$S$_{0.07}$ | $T_c \sim 25$ K for FeSe, $T_c > 31$ K for FeSe$_{0.93}$S$_{0.07}$ | ARPES | 44, 45 |
| Electric field tuning of thin FeSe films on STO | $T_c^{onset} \sim 40$ K | Electrical transport and diamagnetic measurements | 46-48 |

**Table 2. Dependence of $T_c$ with tensile strain in 1-UC FeSe films on different substates with various in-plane lattice constants.** For consistancy, all $T_c$ values are from ARPES results.

| Substrates | In-plane lattice constant | Tensile strain (relative to bulk value 3.765 Å) | $T_c$ |
|---|---|---|---|
| Nb: BaTiO$_3$/KTaO$_3$ (Rotated lattice) | 3.78 Å | 0.4% | ~ 70 K [69] |
| 3-UC Nb: STO/LaAlO$_3$ | 3.79 Å | 0.7% | ~ 55 K [112] |
| 5-UC Nb: STO/LaAlO$_3$ | 3.81 Å | 1.2% | ~ 62 K [112] |
| Nb: STO | 3.91 Å | 3.9% | ~ 65 K [13, 14] |
| Nb: STO/KTaO$_3$ | 3.99 Å | 6.0% | ~ 70 K [18] |
| Nb: BaTiO$_3$/KTaO$_3$ (Unrotated lattice) | 3.99 Å | 6.0% | ~ 75 K [69] |

**Figure 1. Superconductivity in FeSe films on graphitized SiC (0001) substrate revealed by STM.** (a) STM topography of a FeSe film of ~ 30-UC thick. The crystal structure of FeSe is shown in the inset. (b, c) Normalized differential tunneling conductance (d$I$/d$V$) spectra at various temperatures for (b) 8-UC and (c) 2-UC FeSe films. The insets are temperature dependence of zero bias conductance (ZBC). (d) Relation between superconducting transition temperature $T_c$ and film thickness $d$. $T_c$ can be well scaled as $T_c(d) = T_{c0}(1 - d_c / d)$. (a) Reprinted from [62]. (b-d) Reprinted from [61].

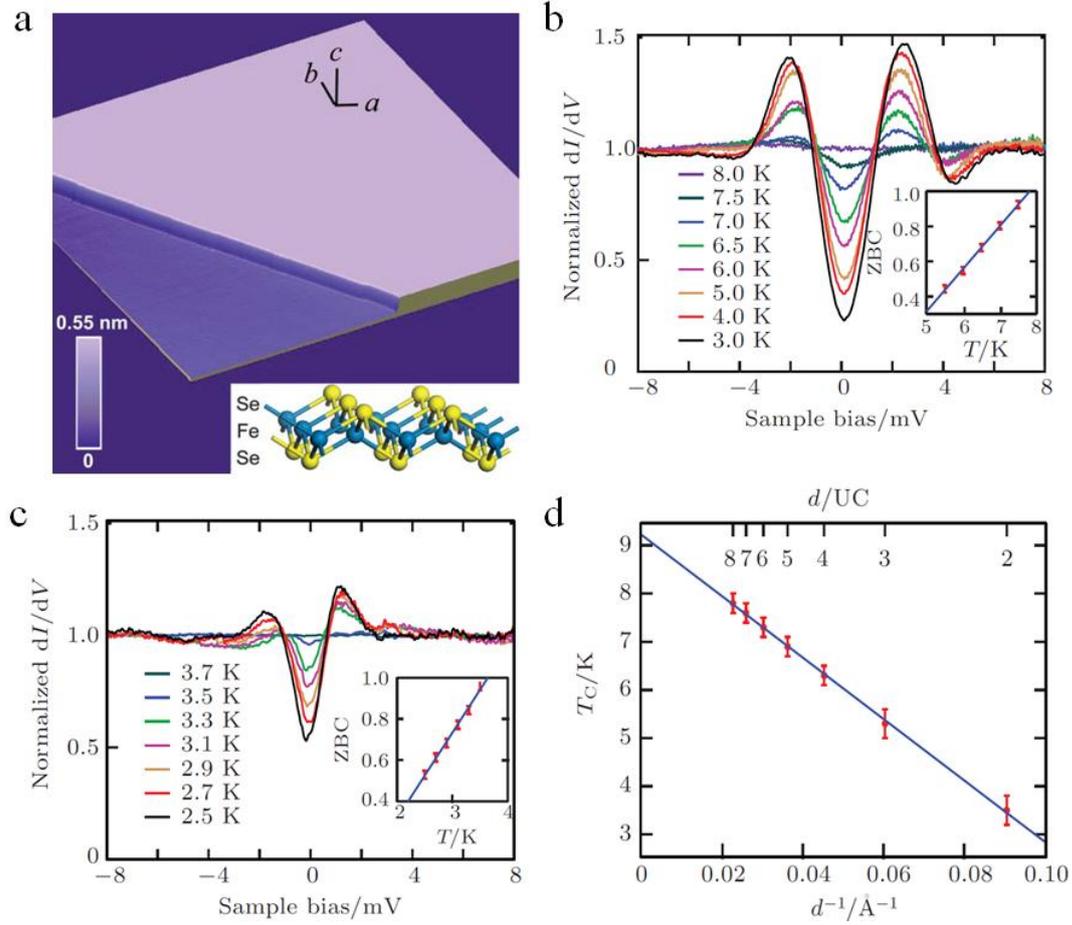

**Figure 2. Detection of high-temperature superconductivity in 1-UC FeSe/Nb: STO (001) by STM.** (a) Schematic structure (side-view) of FeSe films on $TiO_2$-terminated STO substrate along c-axis. (b) STM topography of 1-UC FeSe film. (c) Tunneling spectrum taken at 4.2K reveals the appearance of a superconducting gap. Four pronounced coherence peaks appear at ±20.1 mV and ±9 mV, respectively. (d) Tunneling spectrum taken on 2-UC FeSe film reveals non-superconducting behavior. Reprinted from [10].

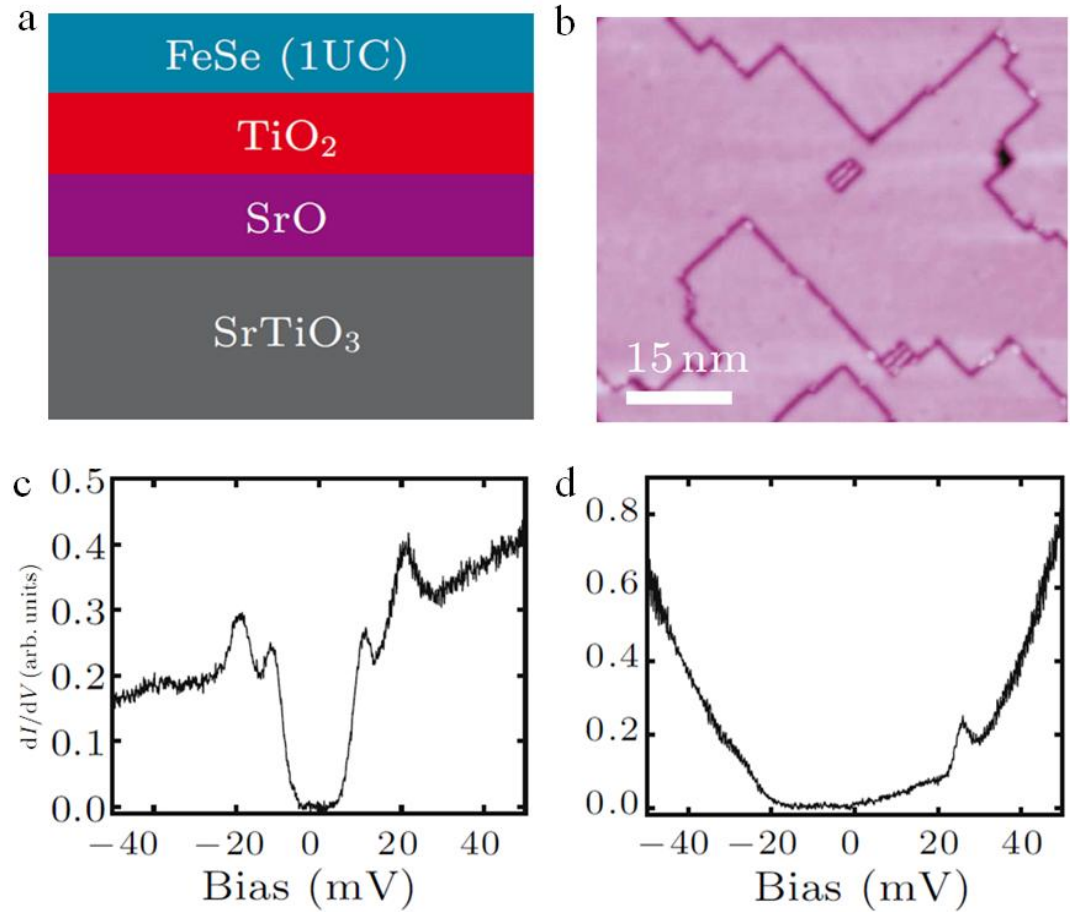

**Figure 3. Direct observation of superconductivity in 1-UC FeSe films on insulating STO (001) by electrical transport and diamagnetic measurements.** (a) Temperature dependence of sheet resistance at magnetic field $B = 0$, indicating $T_c^{onset} \sim 40.2$ K, $T_c^{zero} \sim 23.5$ K. The schematic for transport measurements is shown in the inset. (b) Magnetic susceptibility by commercial magnetic property measurement system (MPMS-SQIUD-VSM) shows a sharp drop ~ 25 K. Inset: typical magnetic hysteresis behavior at 2 K, which confirms the superconducting properties of the 1-UC FeSe film. Reprinted from [11].

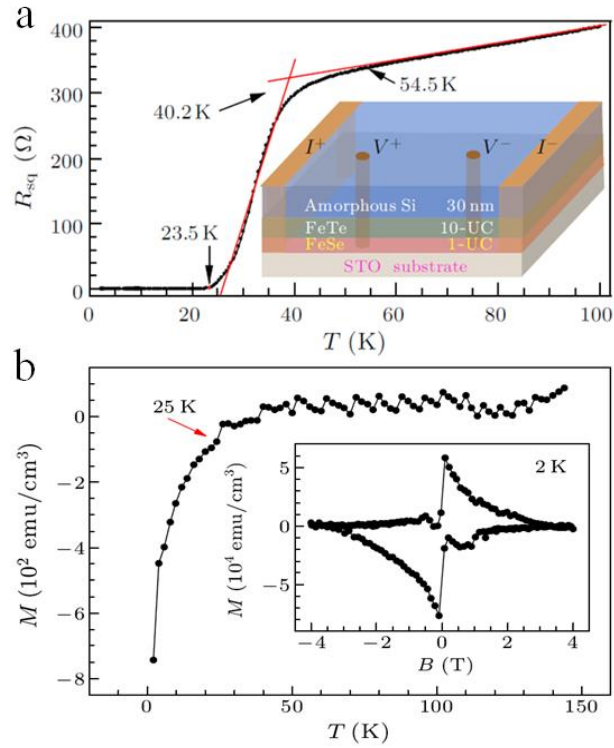

**Figure 4. High critical magnetic field and large critical current density in 1-UC FeSe/STO.** (a, b) Magnetoresistance measured by employing pulsed magnetic field up to 52 T in perpendicular (a) and parallel (b) orientation. (c) $V(I)$ curves measured at temperatures ranging from 2 K to 50 K at magnetic field $B = 0$. (d) Dependence of critical current density ($J_c$) with temperature and perpendicular magnetic field. Reprinted from [11].

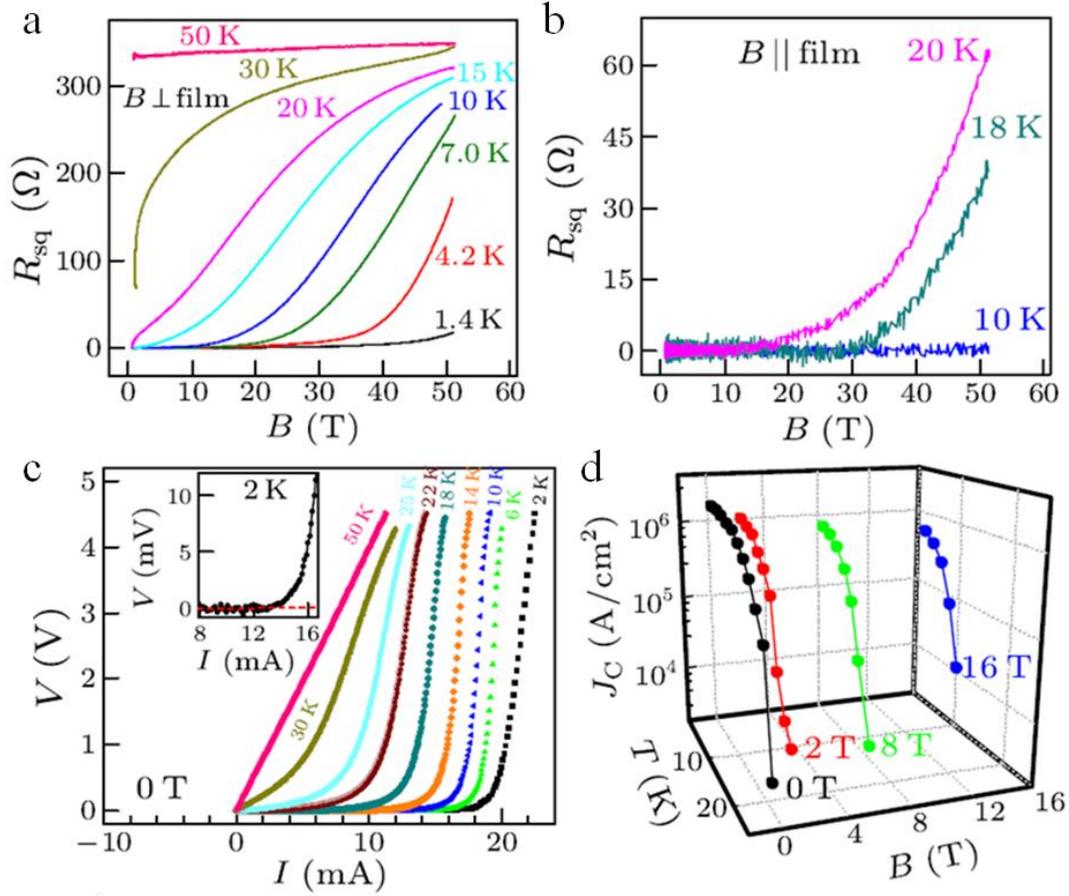

**Figure 5. BKT-like transition of 1-UC FeSe/STO.** (a) $V(I)$ curves at various temperatures plotted in logarithmic scale, the two dashed lines correspond to $V \propto I$ and $V \propto I^3$. (b) Temperature dependence of the power-law exponent $\alpha$ ($V \propto I^\alpha$) indicates BKT-like transition temperature $T_{BKT}$ ~ 23.1 K. (c) $(d\ln(R)/dT)^{-2/3}$ plotted as a function of temperature. The dashed line depicts the expected BKT-like transition with $T_{BKT}$ ~ 23.0 K. Reprinted from [11].

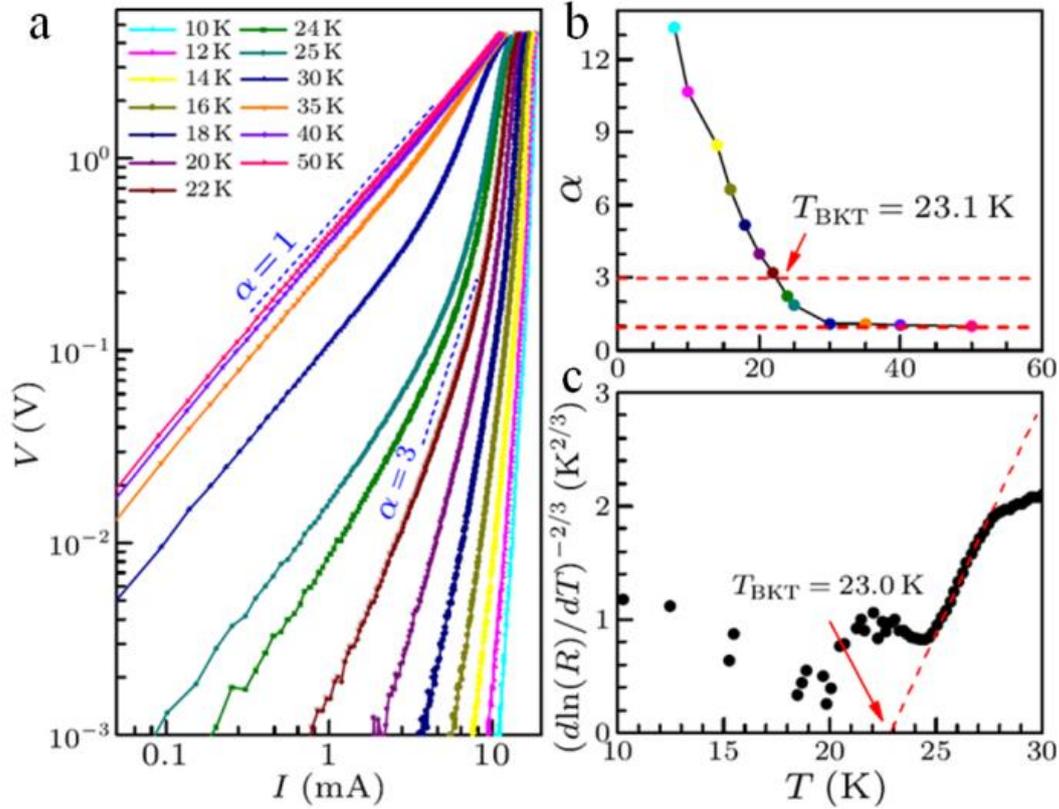

**Figure 6. Analysis of the thermally activated flux flow (TAFF) in 1-UC FeSe/STO.** (a) $R(T)$ curves measured in perpendicular magnetic fields. (b) $\ln(R_{sq})$ vs. $1/T$ at various magnetic fields. Data at low temperatures are fitted by the Arrhenius relation $R(T,H) = R(H)\exp[-U_0(H)/T]$ (solid lines). (c) Field dependence of the activated energy $U_0(H)$. The solid lines show the power-law fittings using $U_0(H) \sim H^{\gamma}$. Reprinted from [19].

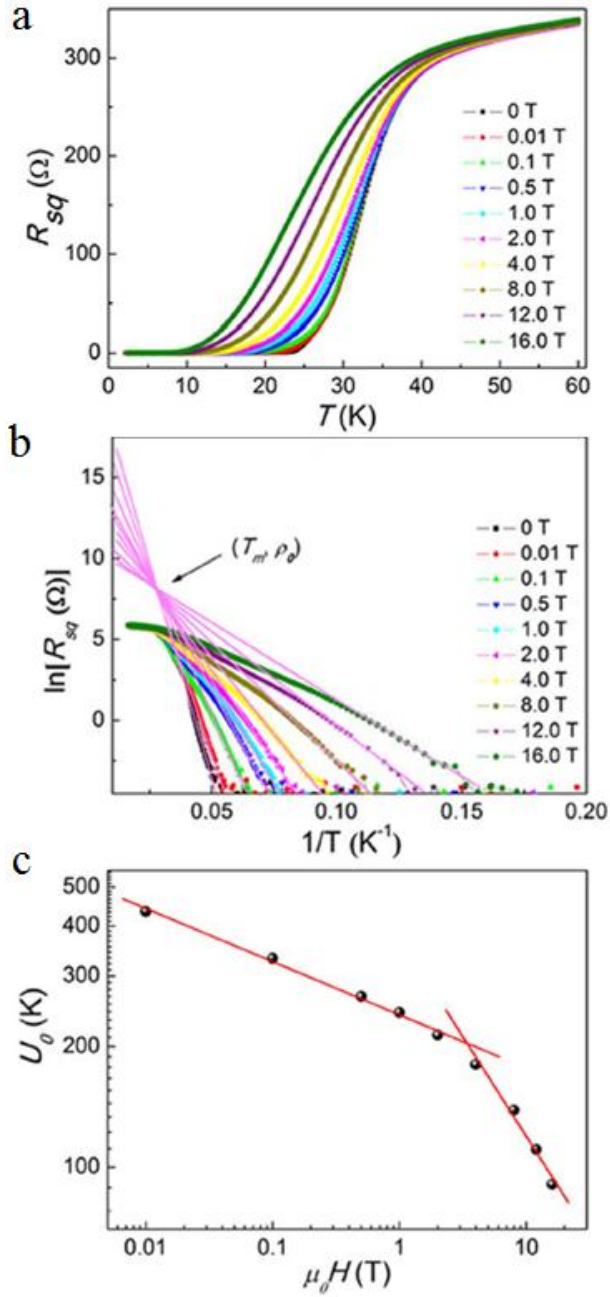

**Figure 7. Hall measurements of 1-UC FeSe/STO.** (a) $R_{xy}(H)$ curves at various temperatures after subtracting the influence of FeTe layers. (b) Hall coefficient and carrier density calculated from data in panel (a). Reprinted from [19].

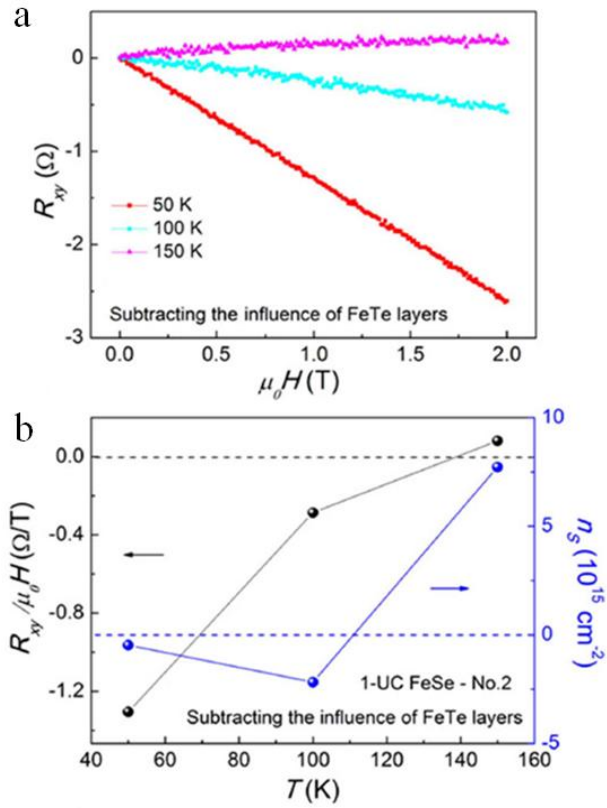

**Figure 8. Dependence of superconducting properties on film thickness and annealing process in FeSe/STO.** (a) $R_{sq}(T)$ curves of FeSe films with different thickness. The annealing time for each film is fixed at 36 h. (b) Thickness dependence of superconducting transition temperature $T_c$. Here, $T_c$ is defined as the crossing temperature of two extended gray lines in panel (a). (c) $R_{sq}(T)$ curves of 5-UC FeSe films with different annealing process (as-grown or after annealing at 500 ℃ for 10 h, 36 h and 55 h, respectively). $T_c$ is labeled beside each curve with the same color. (d) The corresponding temperature-dependent Hall coefficient. Inset: zoom-in of the region near $R_H = 0$. Dominated carrier type transforms from hole to electron while cooling the sample across $T_{e-h}$. Reprinted from [82].

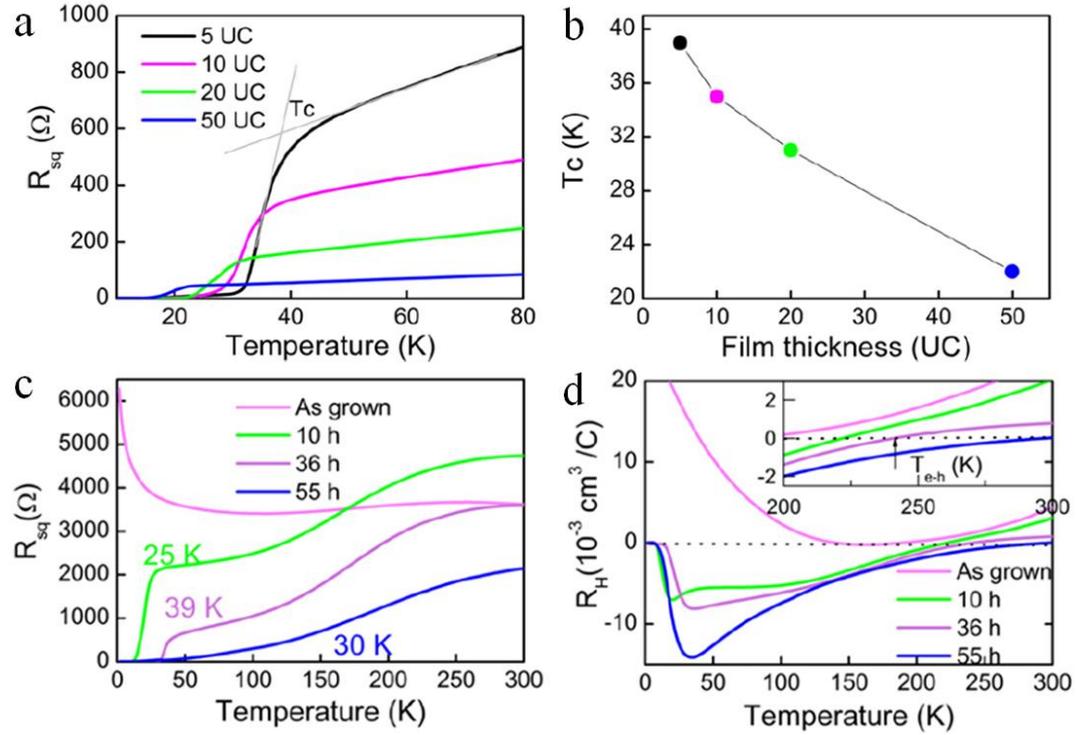

**Figure 9.** *In situ* **special four-tip electrical transport measurement of 1-UC FeSe/Nb: STO**. (a) A schematic for *in situ* transport measurement setup. Resistance is measured by contacting four tips with the sample surface at an inclined angle of 20 °. (b) Temperature dependence of the resistance obtained from *V*(*I*) curves. Above 79 K the sample was cooled by liquid $N_2$. Inset: temperature dependence of resistance taken on a bare STO surface. Reprinted from [20].

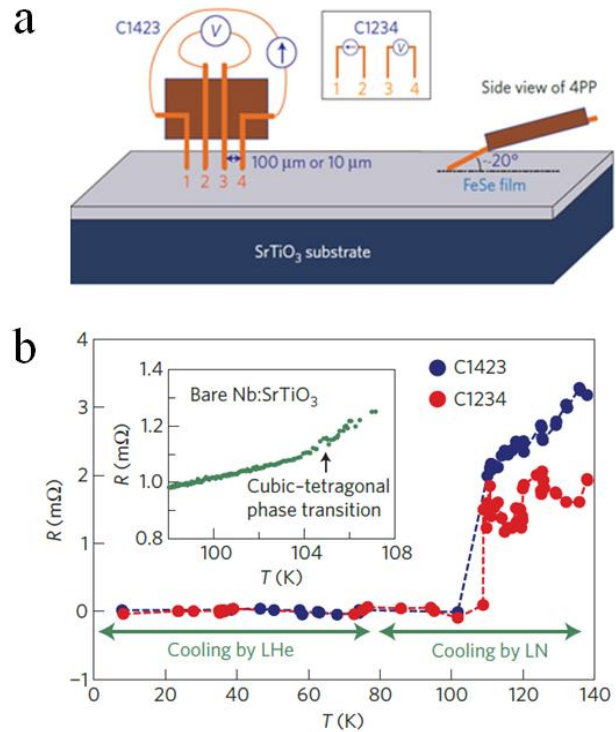

**Figure 10. Magnetic susceptibility (*M*) of 1-UC FeSe/Nb: STO.** Data is measured by commercial magnetic property measurement system (MPMS-SQIUD-VSM). (a) *M*(*T*) curves of 1-UC FeSe/Nb: STO with 10-UC FeTe capping layer measured under 1000 Oe parallel magnetic field. Inset: *M*(*T*) curves of Nb: STO with 10-UC FeTe capping layer measured in same conditions. (b) *M*(*T*) curves of 1-UC FeSe film after subtracting the influence of Nb: STO substrate and FeTe capping layer exhibit a gradual decrease in magnetization starting from ~ 85 K. Reprinted from [19].

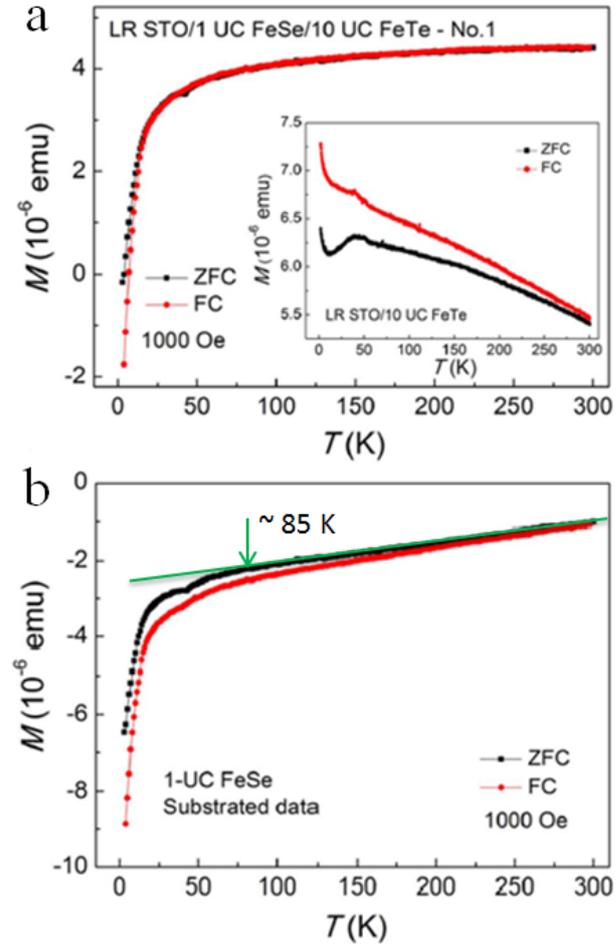

**Figure 11. Evolution of superconducting gap with annealing processes and temperature dependence of superconducting gap in 1-UC FeSe/STO.** (a) Differential tunneling conductance (d$I$/d$V$) spectra taken after annealing at 450, 480, 500, 510, and 530 °C. The corresponding gap sizes are 9.1, 11.8, 12.5, 13.2, and 15.4 meV, respectively. (b) Dependence of superconducting gap with annealing temperature. (c) Normalized d$I$/d$V$ spectra at various temperatures after annealing at 530 °C. (d) Temperature dependence of zero bias conductance (ZBC) gives a extrapolated $T_c$ ~ 68 K. Reprinted from [15].

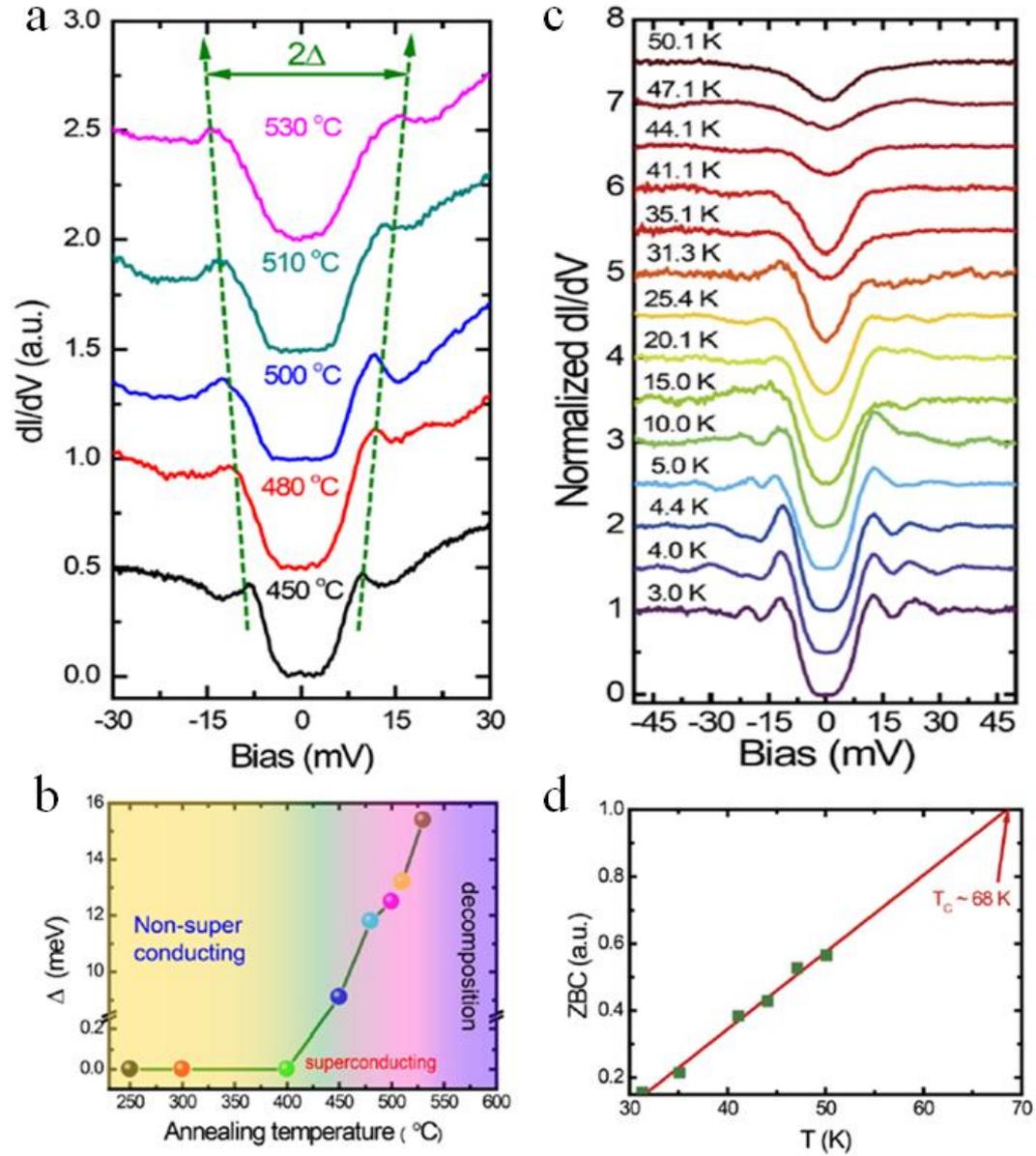

**Figure 12. Quasi-particle interference (QPI) intensity for different scattering channels in 1-UC FeSe/STO.** (a) Schematic of Brillouin zone and Fermi surface. The blue solid ellipse is the electron pocket at M points. Possible scattering vectors ($q_1$, $q_2$, $q_3$) are marked by red arrows. (b) Calculated joint density of states JDOS($q$) according to electronic structure at energies outside the superconducting gap. (c) QPI pattern at bias $V$ = 16.5 meV. The masked areas show the integration windows for different scattering rings. (d) Integrated QPI intensity as a function of energy for three scattering channels in panel (c). (e) Relative difference between QPI patterns at $B$ = 11 T and $B$ = 0 T at bias $V$ = 10.5 meV. (f) Integrated QPI intensity difference between $B$ = 11 T and $B$ = 0 T as a function of energy for three scattering channels in panel (e). Reprinted from [91].

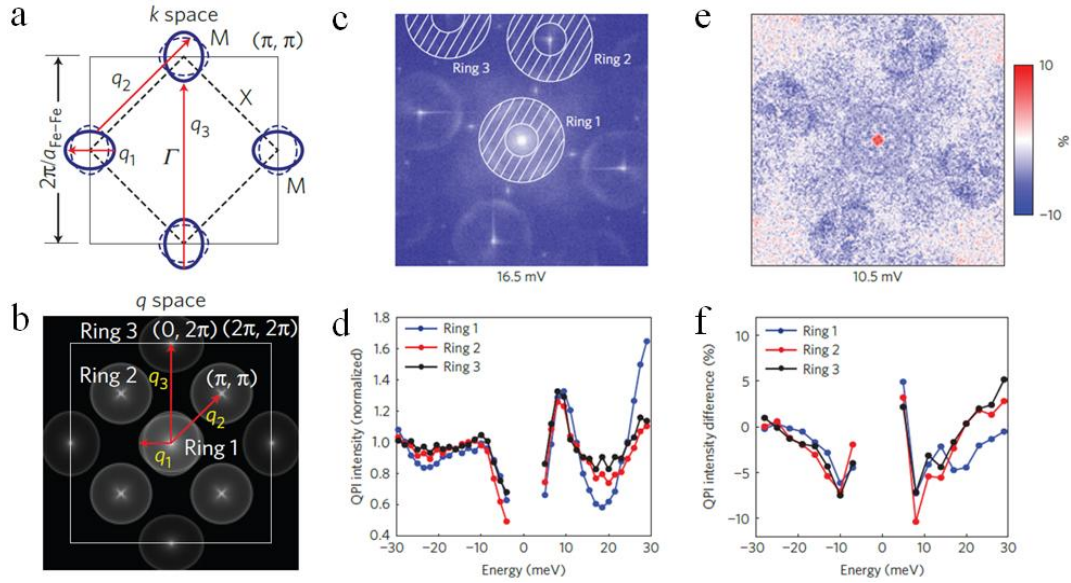

**Figure 13. Local response of superconductivity to magnetic (Cr and Mn) and non-magnetic (Zn, Ag and K) impurities.** (a-e) Topographic images of single adatoms on FeSe/STO films. (f-j) Differential tunneling conductance (d$I$/d$V$) spectra taken along the arrows shown in (a-e). The distance from the measuring points to the center of the atom are marked on the left. In-gap states are induced by magnetic impurities Cr and Mn, but absent on non-magnetic impurities Zn, Ag and K. Reprinted from [91].

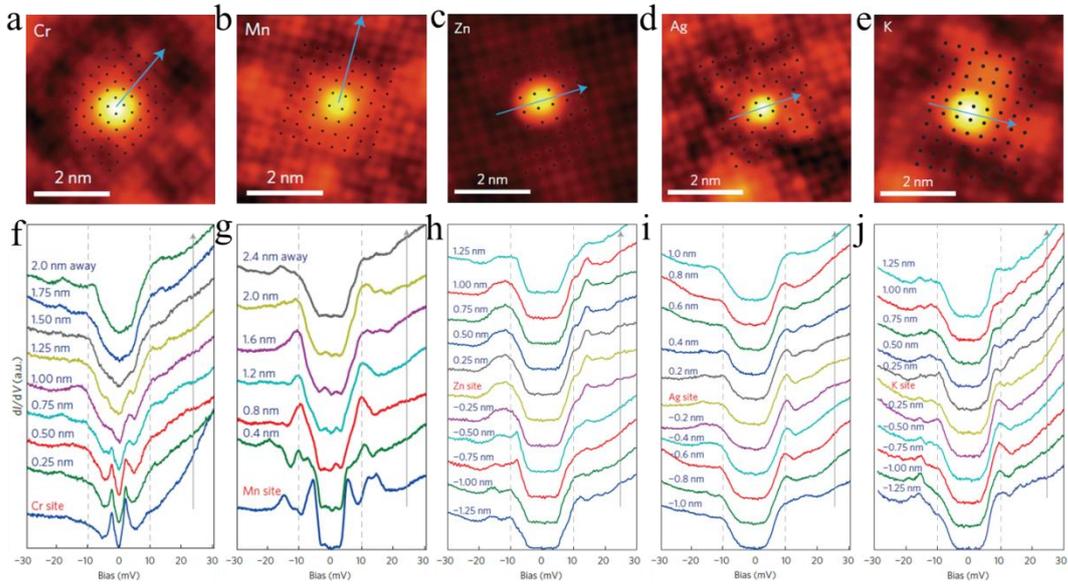

**Figure 14. Fermi surface and band structure of 1-UC FeSe/STO (001).** (a) Fermi surface mapping measured at 20 K consists only of electron pockets at M points. (b) Band structures along cut 1 (left panel) and cut 2 (right panel) at 20 K. The pink dashed lines denoted as $\alpha$, $\beta$ and the purple dashed line denoted as $\gamma$ are guides to the eye, indicative of hole- and electron-like bands, respectively. (c, d) High-resolution Fermi surface mapping and corresponding second derivative image obtained in (c) circular (CR) polarization and (d) linear horizontal (LH) polarization at ~ 120 K. Two overlapping ellipselike electron pockets ($\delta_1$, $\delta_2$) are resolved at M points. (a, b) Reprinted from [12]. (c, d) Reprinted from [109].

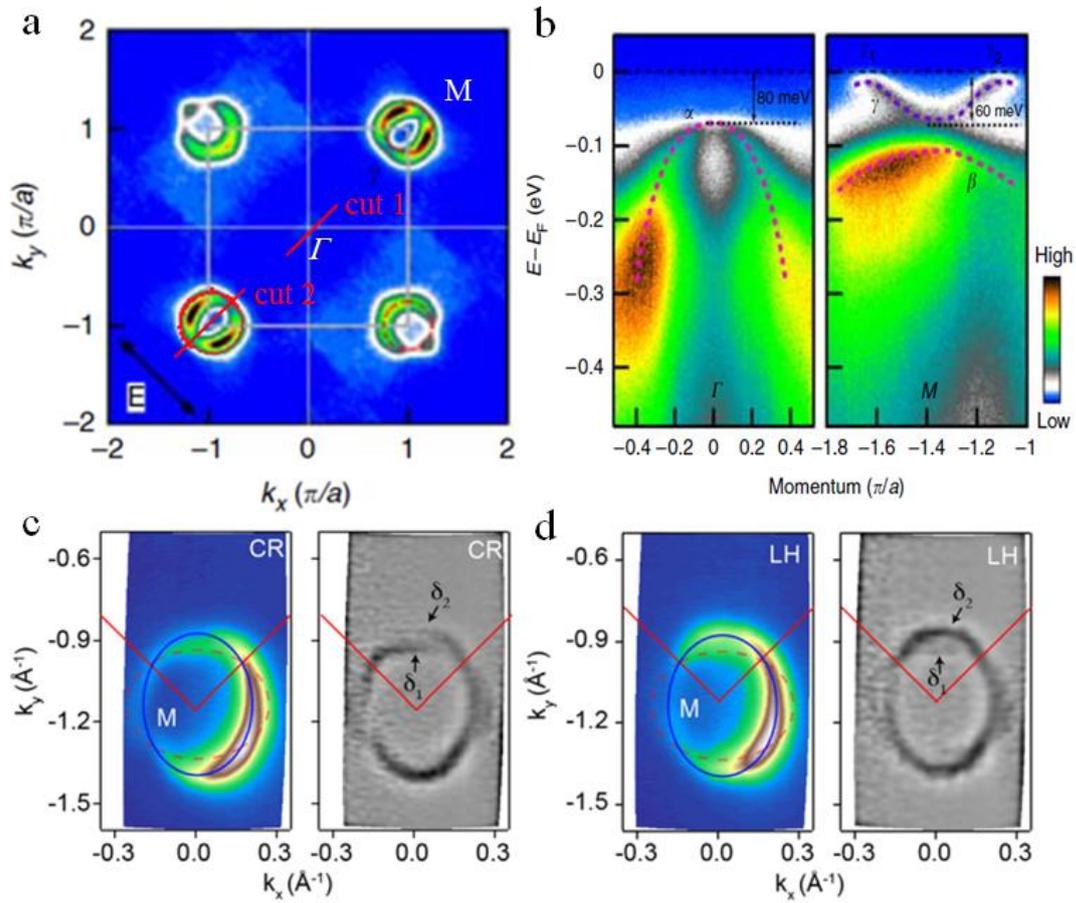

**Figure 15. Temperature and momentum dependence of superconducting gap in 1-UC FeSe/STO (001).** (a) Symmetrized photoemission spectra along the γ Fermi surface measured at different temperatures. (b) Temperature dependence of the superconducting gap. The green line shows the BCS fitting with gap size of ~ 19 meV and $T_c$ ~ 65 K. (c) Superconducting gap distribution in momentum space along the ellipselike electron pocket $\delta_2$ exhibit distinct anisotropy. (a, b) Reprinted from [13]. (c) Reprinted from [109].

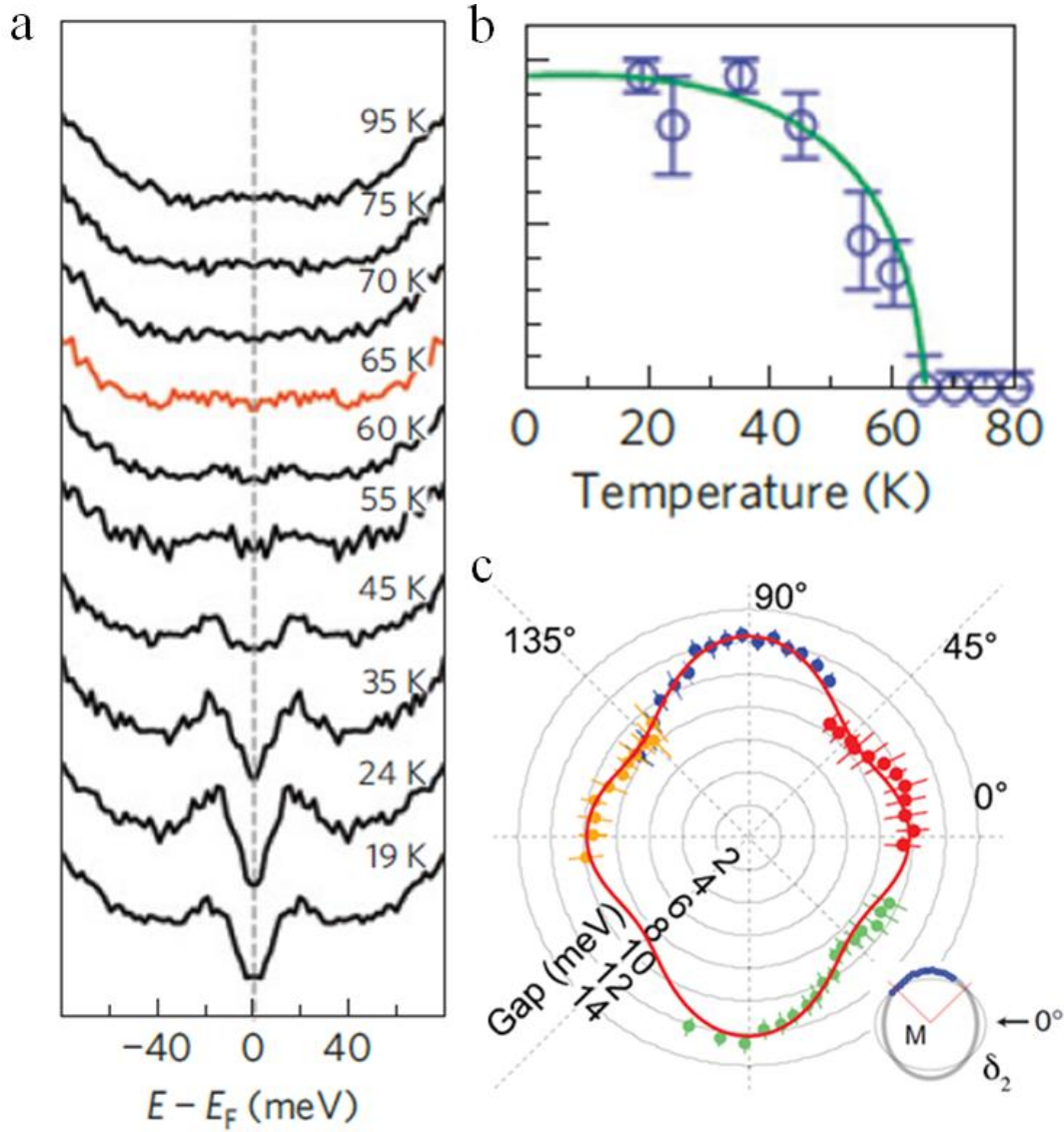

**Figure 16. Electric field induced superconductivity in electrochemically etched ultrathin FeSe films.** (a) Schematic for electron double-layer transistor (EDLT). The gate voltage $V_G$ is applied through the ionic liquid. (b) EDLT can accomplish the function of electrochemical etching and electrostatic carrier doping separately. (c, d) Temperature dependence of normalized sheet resistance $R_s/R_s^{100\,K}$ at $V_G = 5$ V for FeSe films with different film thickness on (c) STO and (d) MgO substrates. The onset superconducting transition temperature $T_c^{on}$ is defined by the intersection of two dashed linear extrapolation lines. (e) $R_s(T)$ curves for a sample with 3.7 nm thick with different gate voltage. The red (purple) solid line represents the fully charged situation at $V_G = 5$ V (fully discharged at $V_G = 0$). $T_c^{on}$ is labeled with triangles. Reprinted from [46].

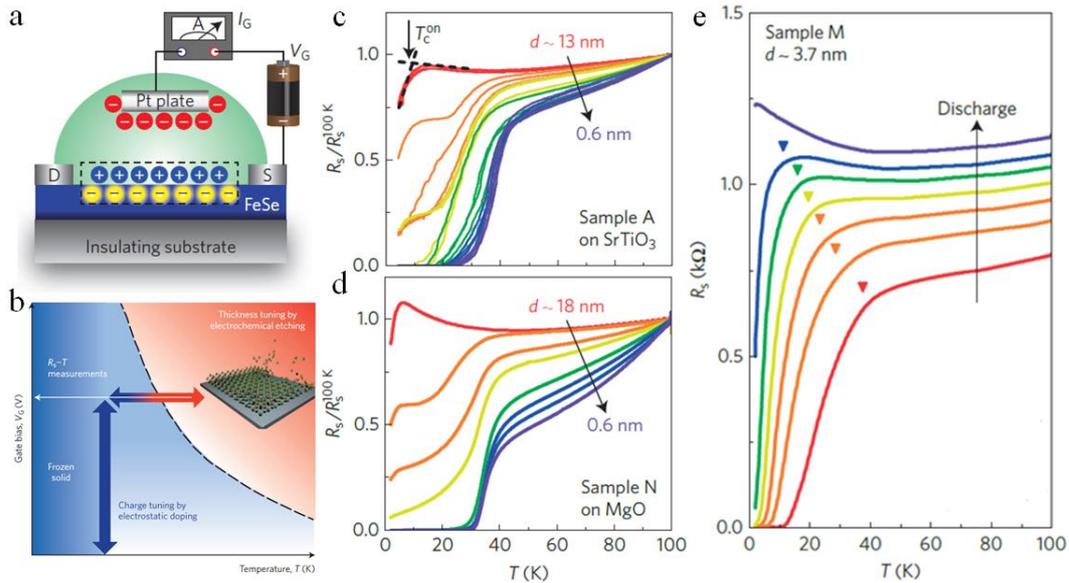

**Figure 17. STM probing in K-coated FeSe films on STO and graphitized SiC substrates.** (a) Dependence of optimized superconducting gap with film thickness for FeSe films on STO (black square) and graphitized SiC (red circle). (b) Comparison of the decay behavior of the superconducting gap (red) and the penetrating field intensity of Fuchs-Kliewer phonon modes in STO (blue) with increasing FeSe layer. (a) Reprinted from [43]. (b) Reprinted from [121].

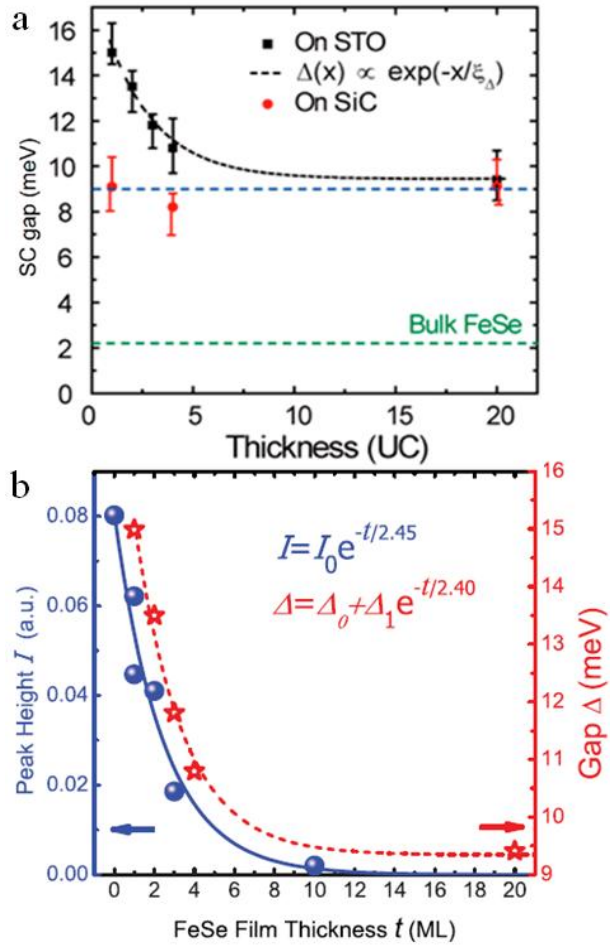

**Figure 18. Theoretical calculation of electron-phonon coupling effects at FeSe/STO interface.**
(a) Frequency-dependent Eliashberg spectral function $\alpha^2F(\omega)$ (solid lines) and electron-phonon coupling constant $\lambda(\omega)$ (dotted lines) of FeSe/STO and bulk FeSe calculated in checkerboard AFM state. (b) $T_c$ enhancement factor $T_c/T_{c0}$ as a function of $\lambda_{inter}$ and $\lambda_{intra}$, where $T_{c0}$ is the superconducting transition temperature in absence of electron-phonon coupling. (a) Reprinted from [125]. (b) Reprinted from [124].

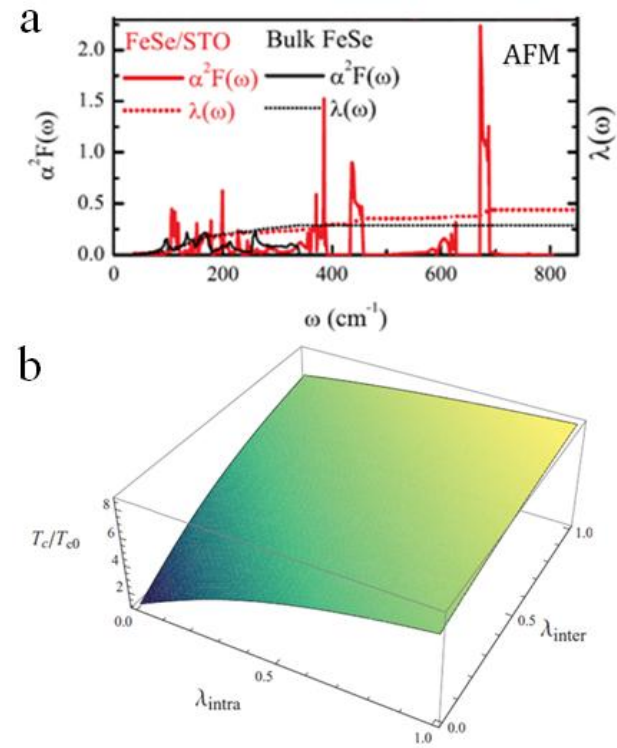

**Figure 19. Experimental observation of replica bands as a signature of interfacial electron-phonon coupling in 1-UC FeSe/Nb: STO (001).** (a) Second derivative image of band structures of the 1-UC film along a high-symmetry cut centered at M point, taken at 16 K. The main bands (A, B and C) and the replica bands (A' and B') are denoted, respectively. (b) EDCs of the original bands of panel (a) at successive momenta along the cut through M point, where the colored squares mark spectra peaks. (c) Replica bands persist in M spectrum (second derivative image) at 90 K for the 1-UC film, which is higher than the gap-opening temperature. (d) Replica bands are absent in M spectrum (second derivative image) for the 2-UC film at 15 K. (e) Original M spectrum for the 1-UC film by high-statistics scan at 10 K. (f) Simulated band structures when assuming both electron and hole bands couple to a 80 meV phonon mode. (g) The calculated EDC intensity shows a good agreement with experimentally measured EDC intensity (after background subtraction). (h) $T_c$ enhancement factor $T_c(v_{eff})/T_c(0)$ as a function of $v_{eff}/J$ for various $J_2/J_1$. Here, $v_{eff}$ is the effective phonon-mediated attraction potential, $J_1$, $J_2$ and $J = \sqrt{J_1^2 + J_2^2}$ represent the nearest, next-nearest-neighbor magnetic exchange constants and antiferromagnetic exchange constant, respectively. Reprinted from [17].

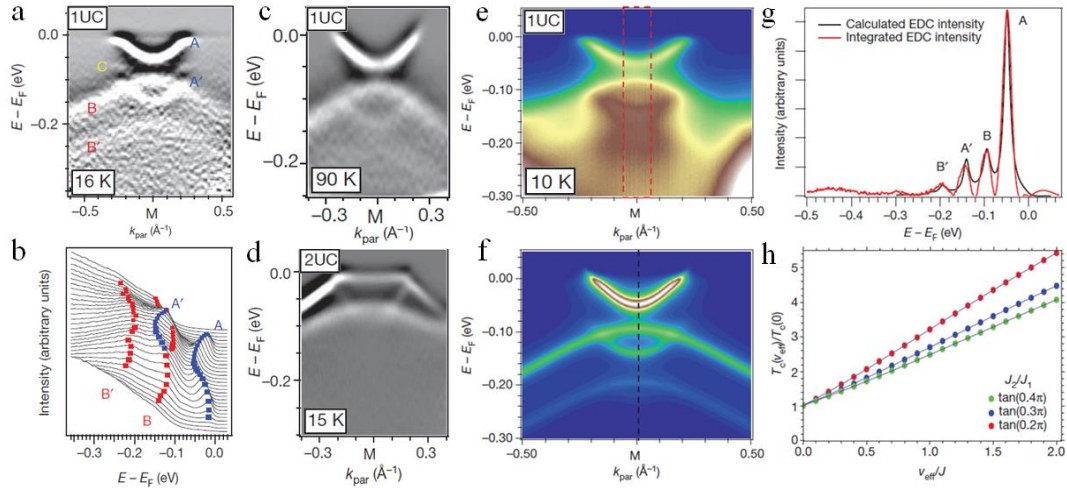

**Figure 20. Comparison of Fermi surface and band structures in several FeSe-based systems.** (a-d) Schematic structure and Fermi surface of 60-UC FeSe/STO (001), K-doped 3-UC FeSe/STO (001), 1-UC FeSe/STO (001) and 1-UC FeSe/TiO$_2$ (100). (e-h) Second derivative images of band structures of the respective systems along the cut through M point. The corresponding $T_c$ values are labelled at the bottom. Reprinted from [122].

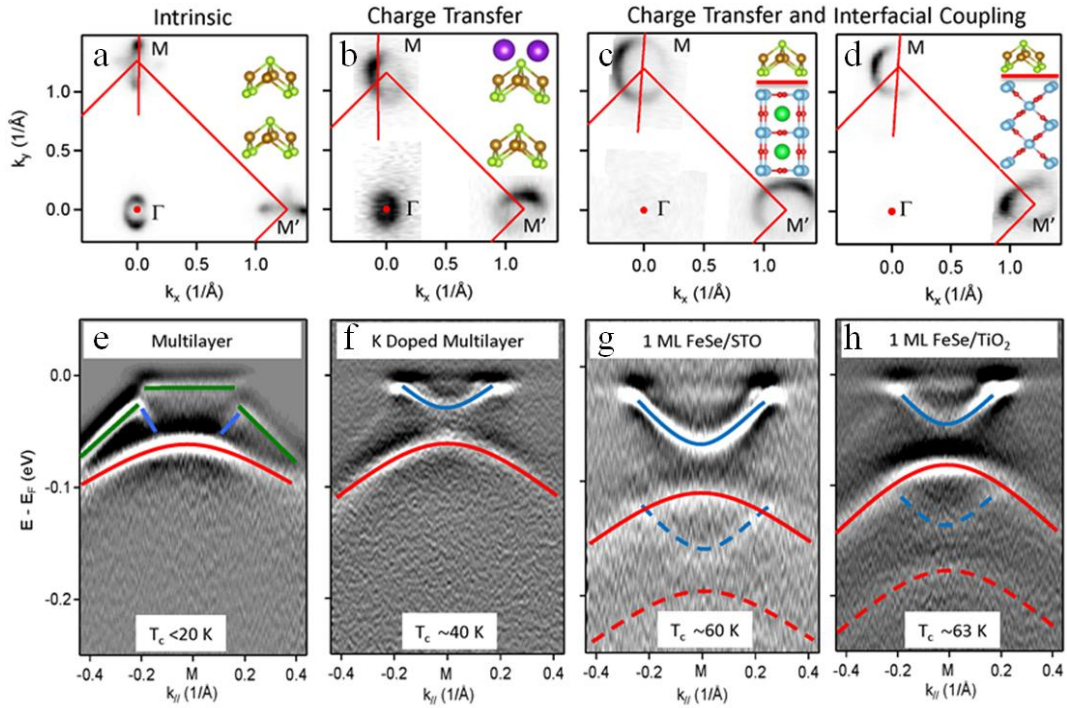